\documentstyle[12pt]{article}                                                   
                                                                                
\title{                                                                         
{\vspace{-3cm} \normalsize                                                      
\hfill \parbox{30mm}{DESY 94-159}}\\[25mm]                                      
Simulating the electroweak phase transition        \\                           
in the SU(2) Higgs model                           \\[8mm]}                     
\author{                                                                        
{Z. Fodor}\thanks{On leave from Institute for Theoretical Physics,              
E\"otv\"os University, Budapest, Hungary.},                                     
J. Hein, K. Jansen, A. Jaster,
{I. Montvay}\thanks{After 1.9.1994 on leave at Theory Division, CERN, 
CH-1211 Geneva 23, Switzerland.} \\[4mm]                              
Deutsches Elektronen-Synchrotron DESY, \\                                       
Notkestr.\,85, D-22603 Hamburg, Germany}                                        
                                                                                
\date{August, 1994}                                                             
                                                                                
\newcommand{\be}{\begin{equation}}                                              
\newcommand{\ee}{\end{equation}}                                                
\newcommand{\half}{\frac{1}{2}}

\begin{document}                                                                
\maketitle                                                                      
                                                                                
\begin{abstract} \normalsize                                                    
Numerical simulations are performed to study the finite temperature             
phase transition in the SU(2) Higgs model on the lattice.                       
In the presently investigated range of the Higgs boson mass,                    
below 50 GeV, the phase transition turns out to be of first order and           
its strength is rapidly decreasing with increasing Higgs boson mass.            
In order to control the systematic errors, we also perform          
studies of scaling violations and of finite volume effects.                             
\end{abstract}                                                                  
\hspace{1cm}                                                                    
                                                                                
%%%%%%%%%%%%%%%%%%%%%%%%%%%%%%%%%%%%%%%%%%%%%%%%%%%%%%%%%%%%%%%%%%%%%%%%        
                                                                                
\section{Introduction}                         \label{sec1}                     
 The masses of elementary particles in the Standard Model are generated         
 via the Higgs mechanism by the non-zero vacuum expectation value of            
 the scalar Higgs field.                                                        
 At high temperatures, above the scale of the vacuum expectation value,         
 the Higgs mechanism is not operative, the symmetry of the vacuum               
 gets restored \cite{KIRLIN}.                                                   
 In fact, in the early universe, according to the big bang cosmology,           
 matter first existed in the symmetry restored phase.                           
 As a consequence of expansion and cooling, a non-zero vacuum                 
 expectation value of the scalar field was developed in the phase                 
 transition between the {\em symmetric phase} at high temperatures and          
 the {\em Higgs phase} at lower temperatures.                                       
 The properties of this {\em electroweak phase transition} might have           
 a substantial influence on the later history of the Universe.                  
 For instance, since the number of baryons is not conserved in                  
 the minimal standard model \cite{THOOFT}, the small baryon asymmetry           
 of the Universe could perhaps be created in non-equilibrium                    
 processes during a strong enough first order electroweak phase                 
 transition \cite{KURUSH,SHAPOSH}.                                              
 This offers the possibility that the baryon asymmetry can be                   
 explained within the minimal standard model.                                   
 The resolution of this question is therefore a major challenge for             
 elementary particle physics.                                                   
                                                                                
 The standard calculational method for the study of the symmetry                
 restoring electroweak phase transition is resummed perturbation theory         
 \cite{CARRIN,BUFOHEWA,ARNESP,FODHEB}.                                          
 In the Higgs phase perturbation theory is expected to work well for            
 not very high Higgs boson masses, since the couplings are small.               
 In the high temperature symmetric phase, however, the situation is             
 similar to high temperature QCD: irreparable infrared singularities            
 occur which prevent a quantitative control of graph resummation                
 \cite{LINDE}.                                                                  
 Since the calculation of physical quantities characterizing the                
 phase transition requires the knowledge of both phases, there is               
 a priory no reason why perturbation theory could provide a quantitative        
 treatment of the electroweak phase transition.                                 
 Indeed, the results of perturbation theory show bad convergence.               
                                                                                
 For a better understanding several non-perturbative methods have also          
 been tried.                                                                    
 For simplicity, fermions and the U(1) gauge field are often omitted.           
 This can be expected on general grounds to be a reasonable first               
 approximation.                                                                 
 In this way one is left with the SU(2) Higgs model describing the              
 interaction of a four-component Higgs scalar field with the SU(2)              
 gauge field.                                                                   
 Possible non-perturbative approaches include a block spin procedure            
 leading to evolution equations for average actions \cite{TEWEREWE},            
 the $\epsilon$-expansion at $4-\epsilon$ spatial dimensions                    
 \cite{ARNYAF} and, of course, numerical simulations.                           
 After pioneering works \cite{DAMHEL,EVJEKA}, recent numerical                  
 simulations concentrated on the understanding of the finite                    
 temperature behaviour of the SU(2) Higgs model at large Higgs                  
 boson masses near and above the W-boson mass \cite{BIKS}.                      
 Another non-perturbative approach is based on dimensional reduction,           
 studying the three-dimensional effective Higgs theory, which is                
 obtained in the high-temperature limit \cite{KARUSH,JAKAPA,FKRSH}.                   
 A further simplification leads to an effective scalar theory                   
 \cite{JANSEU}, which has also been studied numerically in the reduced          
 model \cite{KANEPA}.                                                           
                                                                                
 The non-perturbative investigations of the electroweak phase                   
 transition did not yet lead to a convincing unique picture.                    
 Therefore, we decided to perform a large scale numerical simulation            
 of the symmetry restoring phase transition in the SU(2) Higgs model.           
 We stay in the original four-dimensional theory without reduction.             
 This has the advantage of keeping the number of bare parameters small          
 and not introducing any further approximations beyond the lattice              
 regularization.                                                                
 First results have been published in a recent letter \cite{CFHJJM}.            
 Here we give a detailed description of the techniques used and            
 include additional results.                                                    
 As it is known from previous studies \cite{BIKS}, for Higgs boson              
 masses near and above the W-boson mass the numerical simulations in            
 the original four-dimensional model are technically difficult.                 
 Therefore we restrict the present calculations to smaller Higgs                
 boson masses below 50 GeV.                                                     
 Since this region of parameters of the minimal standard model is               
 already excluded by experiments, our present scope is merely                   
 theoretical because we would like to check the validity of                     
 some other theoretical approximation schemes, e.g.~resummed                   
 perturbation theory.                                                           
 We plan to extend this investigation to heavier Higgs boson masses             
 in future papers.                                                              
                                                                                
%%%%%%%%%%%%%%%%%%%%%%%%%%%%%%%%%%%%%%%%%%%%%%%%%%%%%%%%%%%%%%%%%%%%%%%%        
\subsection{Lattice action}                          \label{sec1.1}             
 The lattice action of the SU(2) Higgs model is conventionally                  
 written as                                                                     
$$                                                                              
S[U,\varphi] = \beta \sum_{pl}                                                  
\left( 1 - \frac{1}{2} {\rm Tr\,} U_{pl} \right)                                
$$                                                                              
\be \label{eq1.1}                                                               
+ \sum_x \left\{ \half{\rm Tr\,}(\varphi_x^+\varphi_x) +                        
\lambda \left[ \half{\rm Tr\,}(\varphi_x^+\varphi_x) - 1 \right]^2 -            
\kappa\sum_{\mu=1}^4                                                            
{\rm Tr\,}(\varphi^+_{x+\hat{\mu}}U_{x\mu}\varphi_x)                            
\right\} \ .                                                                    
\ee                                                                             
 Here $U_{x\mu}$ denotes the SU(2) gauge link variable, $U_{pl}$                
 is the product of four $U$'s around a plaquette and                            
 $\varphi_x$ is a complex $2 \otimes 2$ matrix in isospin space                 
 describing the Higgs scalar field and satisfying                               
\be \label{eq1.2}                                                               
\varphi_x^+ = \tau_2\varphi_x^T\tau_2 \ .                                       
\ee                                                                             
 The bare parameters in the action are $\beta \equiv 4/g^2$ for                 
 the gauge coupling, $\lambda$ for the scalar quartic coupling and              
 $\kappa$ for the scalar hopping parameter related to the bare                  
 mass square $\mu_0^2$ by $\mu_0^2 = (1-2\lambda)\kappa^{-1} - 8$.              
 Throughout this paper we set the lattice spacing to one ($a=1$),               
 therefore all the masses and correlation lengths etc.\ will always be          
 given in lattice units, unless otherwise stated.                               
                                                                                
 In order to fix the physical parameters in a numerical simulation              
 one has to define and compute some suitable renormalized quantities            
 at zero temperature.                                                           
 The physical Higgs mass $M_H$ and W-mass $M_W$ can be extracted from           
 correlation functions of different quantities (see section \ref{sec5}).        
 The renormalized gauge coupling can be determined from the static              
 potential of an external SU(2) charge pair, measured by Wilson loops           
 (see section \ref{sec6}).                                                      
                                                                                
 Since we are interested in the study of the symmetry restoring                 
 phase transition as a function of temperature,                                 
 we use asymmetric lattices: the small temporal extensions                       
 $L_t=2,3,\ldots$ represent the discretized inverse temperature                 
 $L_t = 1/(aT)$.                                                                
 The other three (spatial) extensions of the lattice have to be much            
 larger, for reaching the thermodynamical limit.                                
 In order to fix the physical parameters at the phase transition                
 we have to determine the zero temperature renormalized parameters at           
 the phase transition points for the $L_t=2,3,\ldots$ lattices.                 
 A renormalized gauge coupling near the physical value                          
 $g_R^2 \simeq 0.5$ can be obtained near $\beta=8$ \cite{LAMOWE}.               
 As stated before, we would like to have lighter Higgs boson masses             
 than studied in \cite{BIKS}.                                                   
 Hence for the bare quartic coupling we have chosen values near                 
 $\lambda = 0.0001$ (in this paper referred to as {\em low}) and near           
 $\lambda = 0.0005$ (referred to as {\em high}).                                
 In the present paper the inverse temperature in lattice units will             
 be restricted to $L_t=2$ and $L_t=3$.                                          
 Therefore the indices of the four sets of numerical simulations will be        
\pagebreak\begin{itemize}                                                                 
\item                                                                           
 $l2$ for low $\lambda$ and $L_t=2$;                                            
\item                                                                           
 $l3$ for low $\lambda$ and $L_t=3$;                                            
\item                                                                           
 $h2$ for high $\lambda$ and $L_t=2$;                                           
\item                                                                           
 $h3$ for high $\lambda$ and $L_t=3$.                                           
\end{itemize}                                                                   
                                                                                
 In the next section the numerical simulation methods will be                   
 discussed.                                                                     
 An important tool for the orientation in bare parameter space will             
 be the invariant effective potential introduced in section \ref{sec3}.         
 Then different groups of numerical simulation results will be                  
 discussed: the location of the phase transition points in                      
 section \ref{sec4}, masses and correlation lengths in                          
 section \ref{sec5}, the renormalized gauge coupling and the                    
 renormalization group trajectories in section \ref{sec6}, the latent           
 heat in section \ref{sec7} and finally the interface tension in                
 section \ref{sec8}.                                                            
 The last section is devoted to the discussion of results and to a              
 summary.                                                                       
                                                                                
%%%%%%%%%%%%%%%%%%%%%%%%%%%%%%%%%%%%%%%%%%%%%%%%%%%%%%%%%%%%%%%%%%%%%%%%        
                                                                                
\section{Monte Carlo simulation}               \label{sec2}                     
 In this section some aspects of the applied Monte Carlo simulation             
 techniques are discussed.
 This can be skipped by readers not interested in technical details.
                                                                                
 The simulations have been performed on the Alenia Quadrics computers           
 of DESY.                                                                       
 The Quadrics Q16 is a massive parallel machine with SIMD%                      
 \footnote{single instruction multiple data}                                    
 architecture which consists of 128 processors (nodes).                         
 Depending on the goals and features of the respective simulation,              
 we use different strategies:                                                   
\begin{itemize}                                                                 
\item                                                                           
 A lattice is assigned to each node.                                            
 No time is wasted for the communications between the nodes.                    
 Limitations of memory allow this only for small enough lattice                 
 extensions.                                                                    
\item                                                                           
 The Q16 may be switched to consist of 16 independent $2^3$ tori.               
\item                                                                           
 The whole machine is arranged as a three-dimensional torus.                    
\end{itemize}                                                                   
 In this way the lattice is distributed over 1, 8 or 128 nodes and              
 one obtains 128, 16 or 1 independent data sets from one run,                   
 respectively.                                                                  
 Of course, the lattice extensions have to be multiples of the                  
 corresponding tori.                                                            
                                                                                
 The Quadrics offers 32 bit floating point arithmetics.                         
 This is sufficient for most of our purposes, except for building               
 global averages, when we use a simple variant of software based                
 double precision arithmetics for summation.                                    
                                                                                
%%%%%%%%%%%%%%%%%%%%%%%%%%%%%%%%%%%%%%%%%%%%%%%%%%%%%%%%%%%%%%%%%%%%%%%%        
\subsection{Updating}                                \label{sec2.1}             
 In a Monte Carlo simulation the autocorrelation of subsequent                  
 configurations is one of the main problems one has to deal with.               
 The autocorrelation is usually worse at phase transitions.                     
 This phenomenon is called {\em critical slowing down}.                         
 Unfortunately, the continuum limit has to be performed in
 this region of parameter space (in our case                     
 $\kappa$, $\beta$ and $\lambda$).
                                                                                
 Due to the small $\lambda$-values we use, large fluctuations of the            
 squared Higgs field length $\rho_x^2 \equiv                                    
 \frac{1}{2}{\rm Tr}\vspace{0.1ex}(\varphi^+_x\varphi_x)$                       
 occur in the Higgs phase and this expectation value shows the largest          
 autocorrelation of all investigated quantities.                                
 In order to reduce the autocorrelation, the updating has to propose            
 a big change of $\rho_x$, with a high probability to accept it.                
 This is achieved by an overrelaxation algorithm for $\rho_x$,                  
 developed by two of us \cite{zoltankarl}.                                      
 As usual, this overrelaxation algorithm is non-ergodic, thus we             
 have to mix it with some ergodic updating steps.                               
                                                                                
 A further improvement of the autocorrelation was achieved by the               
 replacement of the ergodic Metropolis step considered in                       
 \cite{zoltankarl} with a heatbath algorithm proposed to us by                  
 Burkhard Bunk \cite{bunkpriv}.                                                 
 This algorithm works on the four real components of the                        
 $\varphi_x$-field.                                                             
 As far as we know, this algorithm has not appeared in a                        
 publication, therefore we would like to describe it in some detail.            
                                                                                
 The idea is to divide the action in two parts, a quadratic one and the         
 rest proportional to $\lambda$.                                                
 In a heatbath step a new $\varphi_x$ is proposed according to the              
 quadratic part of the action and the remaining parts are taken into            
 account by an additional accept-reject procedure.                              
 Starting from the lattice action (\ref{eq1.1}), the                            
 $\varphi_x$-dependent part can be written in the following form:               
\be                                                                             
S(\varphi_x)= (\varphi_{x,0}-b_{x,0})^2+ \dots +%                               
(\varphi_{x,3}-b_{x,3})^2 + \lambda(\rho^2_x-1)^2 + {\rm const.}                
 \ee                                                                            
 Here we used the notations                                                     
\begin{eqnarray}                                                                
\varphi_x &=& \varphi_{x,0}\,{\bf 1} + i
\sum_{m=1}^3 \varphi_{x,m} \,\tau_m \ ,\nonumber  \\                            
b_{x,0} &\equiv&  \frac{\kappa}{2}                                              
\sum_{\nu=1}^4 {\rm Tr\hspace{0.1ex}}%                                          
\left( \varphi^+_{x+\hat\nu} U_{x\nu}%                                          
+ U_{x-\hat\nu,\nu}\varphi_{x-\hat\nu}%                                         
\right) \ , \nonumber\\                                                         
b_{x,m} &\equiv&  \frac{i\kappa}{2}                                             
\sum_{\nu=1}^4 {\rm Tr\hspace{0.1ex}}%                                          
\left( \varphi^+_{x+\hat\nu} U_{x\nu} \tau_m %                                  
+\varphi^+_{x-\hat\nu} U^+_{x-\hat\nu,\nu} \tau_m %                             
\right) \ . \nonumber                                                           
\end{eqnarray}                                                                  
 In this form of the action a term proportional to                              
 $\rho^2_x$ can be shifted from the quadratic part to the quartic               
 part.                                                                          
 We introduce a factor $\zeta_x$ to express this freedom:                       
\be \label{zetaaction}                                                          
 S(\varphi_x)=                                                                  
\zeta_x\left[\sum_{j=0}^3                                                       
\left(\varphi_{x,j}-\frac{1}{\zeta_x}b_{x,j}\right)^2                           
\right]+                                                                        
\lambda\left(\rho^2_x-\frac{1}{2\lambda}                                        
(2\lambda-1+\zeta_x)\right)^2+                                                  
{\rm const.'}                                                                   
\ee                                                                             
 For a good acceptance the minima of quadratic and quartic                      
 parts in (\ref{zetaaction}) should coincide.                                   
 So we get for $\zeta_x$ the equation                                           
\be                                                                             
2 \lambda |b_x|^2 = \zeta^2_x(\zeta_x+2\lambda-1) \ ,\qquad                     
{\mbox{with:}}\qquad |b_x|^2 = \sum_{j=0}^3 b_{x,j}b_{x,j} \ .                  
\ee                                                                             
 This cubic equation cannot be solved fast enough in an updating.               
 Starting from the observation $\zeta_x=1$ for $|b_x|=1$, we split              
 $\zeta_x$ in $\zeta_x=1+\varepsilon_x$ and get the following                   
 approximate expression for the optimal $\zeta_x$:                              
\be                                                                             
\zeta_x = 1-2\lambda+2\lambda\cdot|b_x|^2 +                                     
{\cal O}(\varepsilon_x^2,\varepsilon_x \lambda) \ .                             
\ee                                                                             
 This approximation works very well in a sufficient range of $|b_x|$.           
 In practice the average acceptance of this algorithm turned out to be          
 larger than 98\%.                                                              
                                                                                
 For the SU(2)-variables $U_{x\mu}$ and                                         
 $\alpha_x \equiv {\varphi_x}/{\rho_x}$ we use standard                         
 overrelaxation methods \cite{zoltankarl,wolffschladming}.                      
 Because the above described heatbath algorithm for $\varphi_x$ offers          
 also an ergodic update for the angular part $\alpha_x$ of the scalar           
 field $\varphi_x$, we need an ergodic update for $U_{x\mu}$ only.              
 For this purpose we use the heatbath algorithm described in                    
 \cite{haanfabricius,KENPEN}.                                                   
                                                                                
 In all updatings, the random number generator proposed and implemented         
 by Martin L\"uscher \cite{luscherran} is applied.                              
 It is based on an algorithm of Marsaglia and Zaman \cite{MarZa}.               
 The latter algorithm is known by the name RCARRY, if the parameters            
 have been chosen appropriately.                                                
 RCARRY offers an extrem long period $> 10^{171}$, but                          
 unfortunately it owns some short range correlations.                           
 As it has been shown \cite{luscherran}, on long range, a chaotic               
 nature of the algorithm comes to light.                                        
 Skipping from time to time some hundreds of numbers in the sequence            
 the correlation is practically eliminated.                                     
 Due to the skip this random number generator is relatively slow.               
 In order to benefit from the parallel architecture, the random                 
 number generator has to be initialized independently on the nodes of           
 the machine.                                                                   
                                                                                
 For updating the field configurations a combination of the above               
 described five algorithms is used.                                             
 We choose some basic sequence of elementary updatings for the different        
 sets of field variables, which is repeated periodically many times.            
 The whole sequence, which visits every variable at least once but              
 usually many times, will be called {\em sweep}.                                
 The optimization of this basic sequence making up a sweep is                   
 a difficult but important task, which will be discussed in the                 
 next subsection.                                                               
                                                                                
%%%%%%%%%%%%%%%%%%%%%%%%%%%%%%%%%%%%%%%%%%%%%%%%%%%%%%%%%%%%%%%%%%%%%%%%        
\subsection{Autocorrelations}                        \label{sec2.2}             
 Let us consider the autocorrelation of a quantity                              
 $Q[U_{x\mu},\varphi_x]$ measured on a sequence of field                        
 configurations.                                                                
 $Q_n$ is the value of $Q[U_{x\mu},\varphi_x]$                                  
 measured on the $n$-th configuration and                                       
 $\overline{Q} \equiv \frac{1}{N} \sum_{n=1}^N Q_n$ is the average over         
 the configurations.                                                            
 The autocorrelation function for this quantity is defined as:                  
\be                                                                             
{\Gamma}_Q(t) \equiv                                                            
\lim_{N\rightarrow \infty} \frac{1}{N\!-\!t} \sum_{n=1}^{N-t}                   
\left(Q_{n+t}- \overline{Q}\right)%                                             
\left(Q_{n}- \overline{Q} \right) \ .                                           
\ee                                                                             
 By the use of the integrated autocorrelation time                              
\be\label{taudef}                                                               
\tau_{{\rm int},Q} \equiv                                                       
\frac{1}{2}\sum_{t=-\infty}^\infty                                              
\frac{{\Gamma}_Q(t)}{{\Gamma}_Q(0)} \ ,                                         
\ee                                                                             
 the statistical error can be estimated from the variance:                      
\be                                                                             
\sigma_Q^2 \simeq 2 \tau_{{\rm int},Q} \, \frac{1}{N}                           
(\overline{Q^2} - \overline{Q}^2) \ .                                           
\ee                                                                             

 We investigated the integrated autocorrelation time for four                   
 characteristic quantities:                                                     
 $\rho_x^2$, $\rho^2_{(x+L_t/2)}\rho_x^2$, $U_{pl}$ and the                     
 largest calculated Wilson loop.                                                
 The second quantity characterizes the correlation function of                  
 $\rho^2$ at the largest distance.                                              
 In zero temperature simulations the largest Wilson loop had the                
 size $L_s/2 \otimes L_t/2$, with $L_s$ and $L_t$ denoting the                  
 extensions of the lattice in time and space direction, respectively.           
 At finite temperature only the $1 \otimes 1$ Wilson loop was                   
 considered.                                                                    
 For Wilson loops not every orientation was taken: two sides were               
 always in the direction of the largest lattice extension.                      
 Further on, we refer to the correlation function as {\it Cf} and to            
 the Wilson loop as {\it Wl}.                                                   
 If equation (\ref{taudef}) is evaluated on a finite                            
 sequence of configurations, one has to decide where to truncate the            
 sum over $t$.                                                                  
 As mentioned before, the largest $\tau_{\rm int}$-values were found            
 for $\rho_x^2$, so we truncated at the first zero                              
 of ${\Gamma}_{\rho^2_x}(t)$.                                                   
                                                                                
 On larger lattices we usually had 16 independent configurations
 in the computer.                                                                      
 They were evaluated separately and an estimate for the statistical             
 error of the integrated autocorrelation time was obtained from the             
 variance.                                                                      
%%%%%%%%%%%%%%%%%%%%%%%%%%%%%%%%%%%%%%%%%%%%%%%%%%%%%%%%%%%%%%%%%%%%%%%%        
\begin{figure}                                                                  
\vspace{9.0cm}                                                                  
\includegraphics{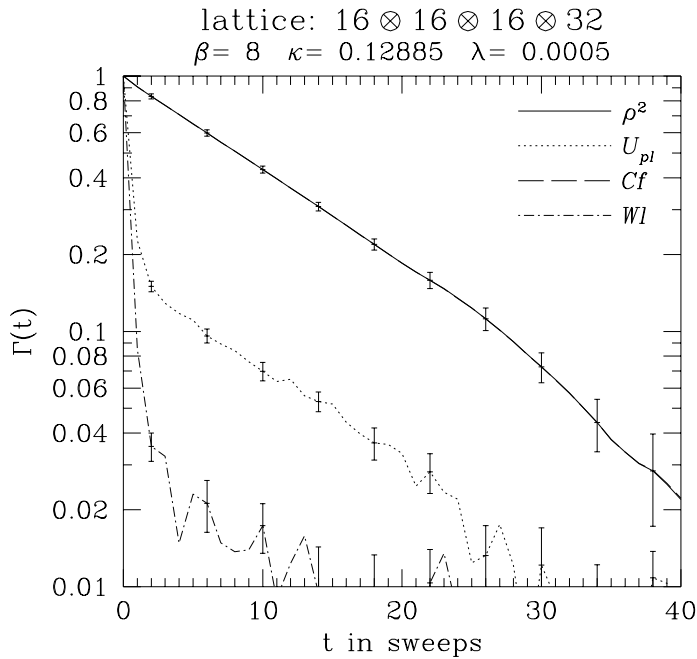}                                              
\begin{center}                                                                  
\parbox{15cm}{\caption{ \label{autot0pic}                                       
 An example of autocorrelation function in the Higgs phase at $T=0$.            
 The curves for $\Gamma_{\rho^2_x}(t)$ and $\Gamma_{C\!f}(t)$
 coincide.                                                   
}}                                                                              
\end{center}                                                                    
\end{figure}                                                                    
%%%%%%%%%%%%%%%%%%%%%%%%%%%%%%%%%%%%%%%%%%%%%%%%%%%%%%%%%%%%%%%%%%%%%%%%        
                                                                                
 We made some investigations how to optimize the autocorrelation by             
 changing the number of calls of the various updatings in the complete          
 sweeps but we did not try to optimize the ordering of the algorithms.          
 Of course, we optimized the autocorrelation in CPU time                        
 since the above changes affect the time requirements of the complete           
 sweeps.                                                                        
 It should be mentioned that the optimal number of calls                        
 depends on lattice size and parameter range.                                   
 The measurement routines were called after each sweep.                
                                                                                
 The autocorrelation function was investigated in the Higgs phase both          
 for $T=0$ and for finite values of $T$.                                        
  In the symmetric phase only finite $T$ was considered.                         
  A typical example of autocorrelation in the Higgs phase is shown by            
 fig.~\ref{autot0pic}.                                                         
%%%%%%%%%%%%%%%%%%%%%%%%%%%%%%%%%%%%%%%%%%%%%%%%%%%%%%%%%%%%%%%%%%%%%%%%        
\begin{figure}                                                                  
\vspace{9.0cm}                                                                  
\includegraphics{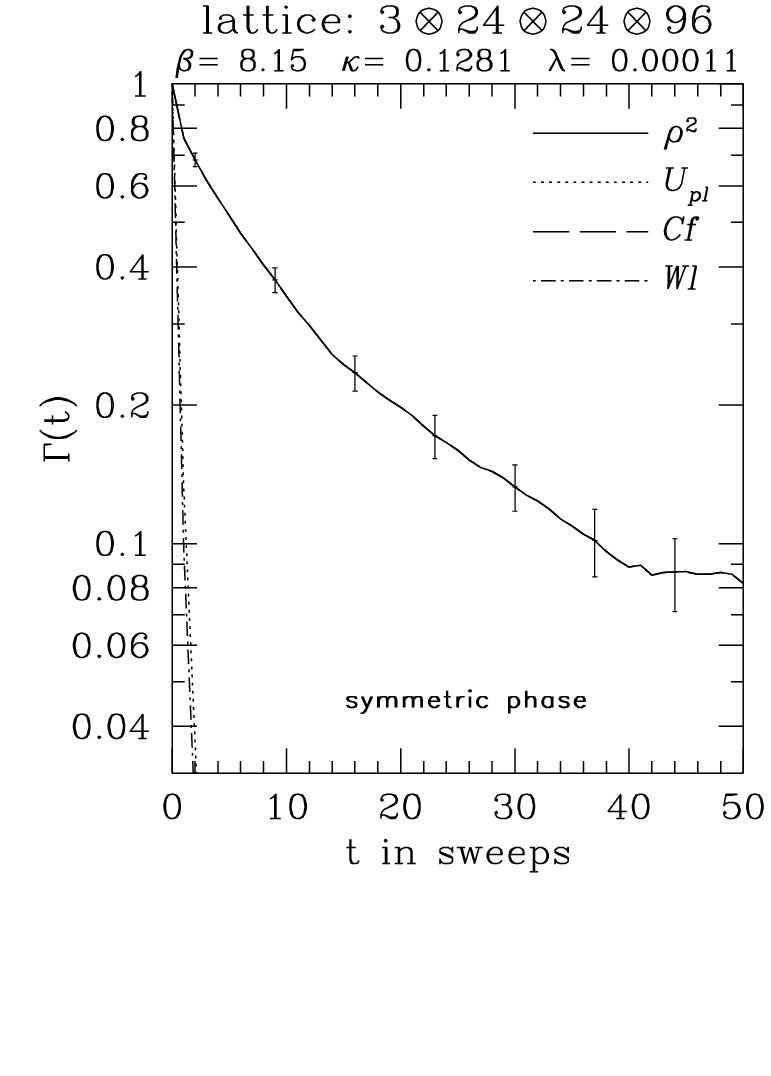}                                           
\includegraphics{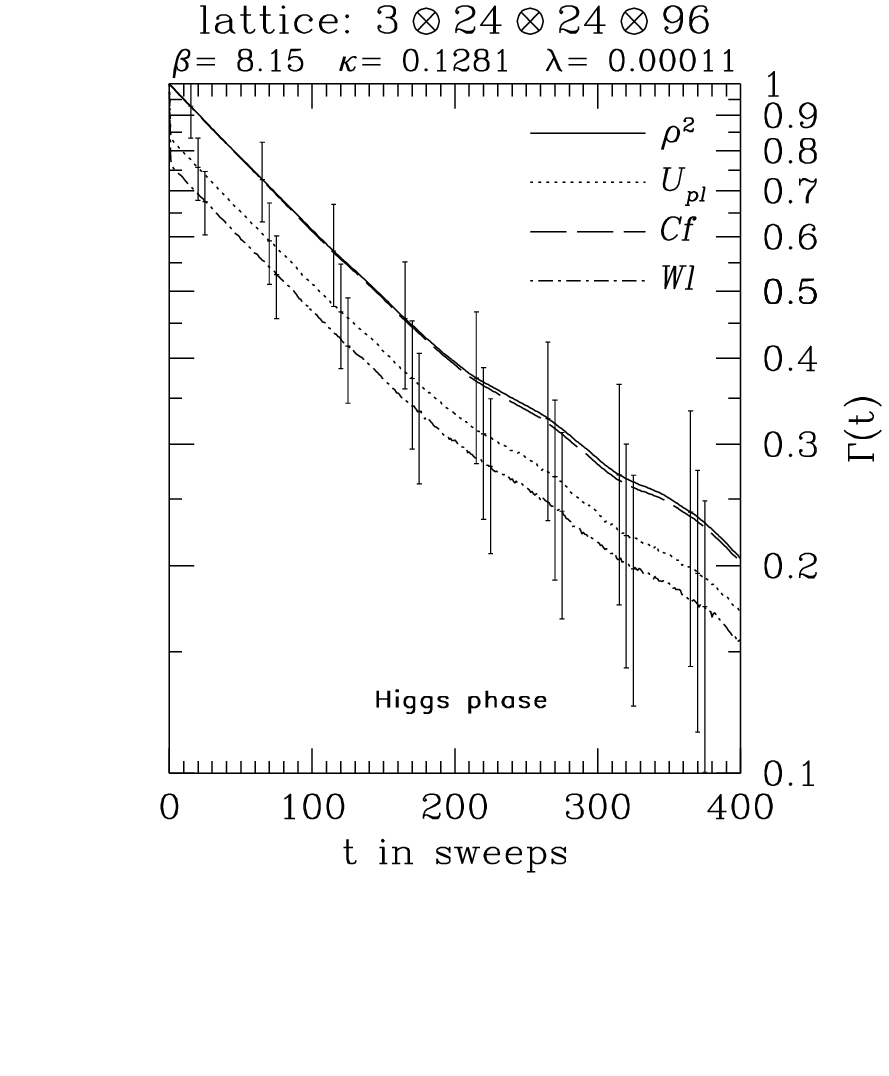}                                           
\begin{center}                                                                  
\parbox{15cm}{\caption{ \label{autocorpic}                                      
 Autocorrelation functions at the phase transition line on a                    
 $3\cdot 24^2\cdot 96$ lattice.                                                 
 The left picture refers to the symmetric phase, the right one to               
 the Higgs phase.                                                               
 Evaluated sweeps: 32000 sweeps in the symmetric phase and             
 80000 sweeps in the Higgs phase.                                      
 In both cases the same updating scheme was applied.                            
 In the symmetric phase $\tau_{{\rm int},\rho_x^2}=21\pm 4$ sweeps and          
 in the Higgs phase $\tau_{{\rm int},\rho_x^2}=217\pm 66$ sweeps.               
}}                                                                              
\end{center}                                                                    
\end{figure}                                                                    
%%%%%%%%%%%%%%%%%%%%%%%%%%%%%%%%%%%%%%%%%%%%%%%%%%%%%%%%%%%%%%%%%%%%%%%%        
%%%%%%%%%%%%%%%%%%%%%%%%%%%%%%%%%%%%%%%%%%%%%%%%%%%%%%%%%%%%%%%%%%%%%%%%        
\begin{table}                                                                   
\begin{center}                                                                  
\parbox{15cm}{\caption{ \label{tab4vergl}                                       
 Integrated autocorrelation time $\tau_{{\rm int},\rho_x^2}$                    
 of $\rho^2_x$ for 5 different updating schemes.                                
 The lattice size is $16^3 \cdot 32$, the parameters are                        
 $\beta=8$ and $\lambda=0.0005$.                                                
 The first item is at $\kappa = 0.12885$ and all the others are at              
 $\kappa = 0.1289$.                                                             
 For each item more than $32000$ sweeps were evaluated.                
}}                                                                              
\end{center}                                                                    
\begin{center}                                                                  
\begin{tabular}{|c|c||c|c|c||c|c|}                                              
\hline                                                                          
\multicolumn{2}{|c||}{heatbath}&                                                
\multicolumn{3}{|c||}{overrelaxation}&                                          
\multicolumn{2}{|c|}{$\tau_{{\rm int},\rho_x^2}$}\\                             
$U_{x\mu}$&$\varphi_x$&$U_{x\mu}$&$\alpha_x$&$\rho_x$&in sec.~of CPU&           
\hspace{1em} in sweeps $\;\;$   \\ \hline \hline                                
1&4&3&3&1&$24.2\pm0.8$&$11.4\pm 0.4$\\ \hline                                   
3&3&3&3&1&$33.3\pm2.3$&$11.4\pm 0.8$\\ \hline                                   
1&1&3&3&1&$34.1\pm2.3$&$19.6\pm 1.3$\\ \hline                                   
1&1&6&6&1&$60.5\pm6.2$&$22.7\pm 2.3$\\ \hline                                   
1&1&3&3&3&$62.1\pm4.7$&$28.0\pm 2.1$\\ \hline                                   
\end{tabular}                                                                   
\end{center}                                                                    
\end{table}                                                                     
%%%%%%%%%%%%%%%%%%%%%%%%%%%%%%%%%%%%%%%%%%%%%%%%%%%%%%%%%%%%%%%%%%%%%%%%        
 The autocorrelation function $\Gamma_{\rho_x^2}(t)$ is to a good               
 approximation a single exponential.                                            
 There is no significant difference between                                     
 $\Gamma_{C\!f}(t)$ and $\Gamma_{\rho_x^2}(t)$.                                 
 The autocorrelation functions for quantities depending only on                 
 $U_{x\mu}$ show a fast fall off for $t$ values very                            
 small compared to $\tau_{{\rm int},\rho^2}$.                                   
 For larger values of $t$ the exponential descent of                            
 $\Gamma_{U_{pl}}(t)$ and $\Gamma_{Wl}(t)$ is the same as of                    
 $\Gamma_{\rho_x^2}$.                                                           
 We always found the initial fall off to                                        
 be larger for $\Gamma_{Wl}(t)$ than for $\Gamma_{U_{pl}}(t)$.                  
                                                                                
 A comparison of autocorrelations in the two phases is given in 
 fig.~\ref{autocorpic}.                                                              
 In the symmetric phase at finite $T$ we found the autocorrelation time         
 for the quantities $U_{pl}$ and $W\!l$ to be less than 1.                      
 The left hand side of figure \ref{autocorpic} displays the extremely           
 fast descent of these autocorrelation functions.                               
 The behaviour of $\Gamma_{\rho^2_x}(t)$ differs from the behaviour             
 in the Higgs phase, because there is a strong curvature in the                 
 logarithmic plot.                                                              
 This could be a signal for a dense spectrum of states.                         
 The integrated autocorrelation time in the symmetric phase is usually           
 much smaller than the one in the Higgs phase at the same parameters            
 and lattice extensions.                                                        
                                                                                
 On a $16^3 \cdot 32$ lattice with parameters                                   
 $\beta=8$, $\kappa=0.1289$ and $\lambda=0.0005$ we                             
 compared four different compositions of the complete sweep.                    
 These parameters give a point in the Higgs phase.                              
 The results for the largest autocorrelation time                               
 $\tau_{{\rm int},\rho_x^2}$ are given in table \ref{tab4vergl}                 
 in sweeps and in CPU seconds of Q16, assuming that the lattice is              
 distributed on the whole machine.                                              
 A comparison of the third and fourth rows shows that more work on              
 the SU(2) variables has no influence on the autocorrelation:                   
 the autocorrelation time in sweeps is about the same for              
 both updating schemes.                                                         
 The fact that more overrelaxation for $\rho_x$                                 
 does not lead to a better autocorrelation is plausible                         
 \cite{zoltankarl}.                                                             
                                                                                
 The autocorrelation measured in sweeps decreases significantly if more         
 heatbath is called but, due to the time needed for the heatbath                
 algorithms, there is no significant difference between the second and          
 the third row of table \ref{tab4vergl}, if the autocorrelation time is         
 measured in CPU time.                                                          
                                                                                
 The updating scheme in the first row of table \ref{tab4vergl} is the           
 best combination we found.                                                     
 It was therefore used in many points.                                          
 Because of the large autocorrelations in the Higgs phase at finite             
 temperatures, which were typically about 10 times longer in sweeps             
 than at $T=0$, we could not compare different updating schemes                 
 there.                                                                         
 Another comparison of updating schemes was performed on a                      
 $18^3 \cdot 36$ lattice with parameters $\beta=8.15$, $\kappa=0.1281$          
 and $\lambda = 0.00011$.                                                       
 The conclusions were very similar.                                             
                                                                                
%%%%%%%%%%%%%%%%%%%%%%%%%%%%%%%%%%%%%%%%%%%%%%%%%%%%%%%%%%%%%%%%%%%%%%%%        
\subsection{Multicanonical simulation}               \label{sec2.3}             
 An important problem of Monte Carlo simulations of a system with first         
 order phase transition is the {\em supercritical slowing down}.                      
 At the transition point the tunneling rate between the two phases is           
 exponentially suppressed for any local update algorithm
 (e.g.~overrelaxation, heatbath).                                                     
 To overcome this problem the multicanonical algorithm was developed            
 \cite{BERG}.                                                                   
 The basic idea is an enhancement of the mixed states, which are                
 suppressed due to the additional free energy of the interfaces.                
 This enhancement is reached by an extra term in the action,                
 i.e.\ $S \rightarrow S + f(O)$.                                                
 This term can be a function of any order parameter $O$.                        
 The easiest way is to use the action and a                                     
 continuous function $f(S)=\beta_k S+\alpha_k$                                  
 with constant $\beta_k$, $\alpha_k$ for $S$                                    
 in the interval $I_k=(S^k,S^{k+1}]$.                                           
 In fact, instead of the lattice action in eq.~(\ref{eq1.1}), we used          
 as an order parameter the modified action                                      
\begin{eqnarray}\label{slog}                                                    
S_{log} \equiv S[U,\varphi]                                                     
-3 \sum_x \log(\rho_x) \ ,                                                      
\end{eqnarray}                                                                  
 which is natural to take if $\rho_x$ is used as an                             
 integration variable in the path integral.                                     
 This choice is particularly convenient, since all the overrelaxation           
 algorithms ($\rho_x$, $\alpha_x$, $U_{x\mu}$) can be used without              
 changes.                                                                       
 The intervals $I_k$ and the parameters $\alpha_k$ and                          
 $\beta_k$ are chosen in such a way that the multicanonical                     
 probability distribution $P_L^{mc}$ is nearly flat.                            
 This is achieved if $f(S_{log}) \approx \log(P_L)$                             
 between the two maxima and is constant elsewhere.                              
 Here $P_L$ is the canonical probability distribution of the action             
 $S_{log}$.                                                                     
 The distribution $P_L$ is obtained in a multicanonical simulation              
 by reweighting $P_L^{mc}$ with $\exp(\beta_k S_{log}+\alpha_k)$.               
                                                                                
 In practice a first choice for the multicanonical parameters is made           
 and they are optimized afterwards.                                             
 If necessary, the procedure is repeated until $P_L^{mc}$ becomes flat.         
 A first guess can be obtained from smaller lattices.                           
 In this way the distribution of the action $S_{log}$ and the link              
 variable $L_\varphi$, defined in eq.~(\ref{eq5.2}),
 was measured on $2\cdot4^2\cdot64$ and                    
 $2\cdot4^2\cdot128$ lattices at the ``low'' value of the quartic               
 coupling ($\lambda=0.0001$).                                                   
                                                                                
 For larger lattices two problems arise.                                        
 The parameters have to be tuned very precisely and the                         
 autocorrelation times become even for optimally tuned values                   
 very large, of the order of ${\cal O}(10000)$ sweeps.                          
 To solve these problems we combined the multicanonical method with the         
 constrained simulation method \cite{BHANOT}.                                   
 In what follows we call this way of simulation                                 
 {\em constrained-multicanonical method}.                                       
                                                                                
 We divide the interval between the two maxima of $P_L$ into                    
 subintervals.                                                                  
 These are chosen to have an overlap with their neighbours.                     
 Starting in one phase we tune the multicanonical parameters such that          
 $P_L^{mc}$ is flat in a given subinterval and suppressed elsewhere.            
 This means that $f(S_{log})$ is approximately equal to $\log(P_L)$             
 in this subinterval and increases rapidly beyond the boundaries.               
 By moving the subinterval one is going from one maximum to the other.          
 At the end every set is reweighted.                                            
 In case of large overlaps between neighbouring intervals                       
 the absolute normalization can be obtained with small errors.                  
                                                                                
 To ensure that this method yields the same result as the pure                  
 multicanonical one, we performed simulations using both                        
 methods on $2\cdot4^2\cdot128$ lattice.                                        
 The results coincide within statistical errors.                                
 Only the constrained-multicanonical algorithm has been used for                
 simulations on $2\cdot8^2\cdot128$ lattice.                                    
                                                                                
 The technical realization of the multicanonical approach by the                
 Metropolis algorithm is straightforward.                                       
 As it has been emphasized above, due to our special choice                     
 of the order parameter, the overrelaxation algorithms can be                   
 used without changes.                                                          
 The modifications for the heatbath algorithms are more involved               
 (see e.g.~\cite{GROSSMANN}).                                                  
 A description of our implementation of heatbath algorithms for                 
 $U_{x\mu}$ and $\varphi_x$ is given in the appendix.                           
 Since the heatbath algorithms are more efficient, we always used them          
 instead of the Metropolis algorithms.                                          
                                                                                
 In our simulations the acceptance rate for the multicanonical heatbath         
 algorithms was very good: for the modified gauge algorithm at least            
 $99\%$ and for the $\varphi$-algorithm at least $96\%$.                        
 The overlaps for the neighbouring intervals  were chosen to be                 
 approximately $40\%$.                                                          
 The number of subintervals was $5$ for the $2\cdot4^2\cdot128$                 
 lattice and $13$ for the $2\cdot8^2\cdot128$ lattice.                          
 The autocorrelation times were on average about $500$ sweeps                   
 for the constrained-multicanonical simulations.                                
 For the two smaller lattices we measured about $2000$ and                      
 $7000$ sweeps as autocorrelation times with the pure multicanonical            
 algorithm.                                                                     
                                                                                
 The easiest way to parallelize the multicanonical algorithm                    
 on the Quadrics Q16 machine is to simulate several lattices                    
 independently on each node.                                                    
 For any other implementation there is a need for communication                 
 between the different nodes for each updating step, since $f$ is a             
 function of the global action.                                                 
 Another disadvantage of partitioning the lattice would be a decrease           
 in the acceptance rate due to simultaneous change of several                   
 variables.                                                                     
                                                                                
%%%%%%%%%%%%%%%%%%%%%%%%%%%%%%%%%%%%%%%%%%%%%%%%%%%%%%%%%%%%%%%%%%%%%%%%        
                                                                                
\section{Invariant effective potential}        \label{sec3}                     
 In the perturbative approach to the electroweak phase transition               
 the most important quantity to compute is the effective potential.             
 Interesting physical observables like latent heat,                             
 surface tension or masses can be extracted from it.                            
 Of course, these latter quantities can also be obtained from                   
 the non-perturbative approach of numerical lattice                             
 simulations by measuring suitable observables.                                 
 Nevertheless, a direct comparison of the effective potential itself            
 from both methods would obviously be desirable.                                
 On the lattice, however, the action is gauge invariant and                     
 so are the observables, as demanded by Elitzur's theorem.                      
 In perturbation theory the effective potential is                              
 calculated in different gauges.                                                      
 (The most popular one is the Landau gauge.)
 In order to compare this with lattice results                                  
 one ought to fix the gauge on the lattice, a notoriously difficult             
 task in particular for non-abelian gauge groups.                               
                                                                                
 A way out is the study of the gauge invariant effective                        
 potential that has been initiated recently \cite{LUE,BUFOHE}                   
%%%%%%%%%%%%%%%%%%%%%%%%%%%%%%%%%%%%%%%%%%%%%%%%%%%%%%%%%%%%%%%%%%%%%%%%        
\footnote{Another possibility could be                                          
the effective potential computed from the lowest eigenvalue of the              
Laplace operator\cite{PA}.}.                                                    
%%%%%%%%%%%%%%%%%%%%%%%%%%%%%%%%%%%%%%%%%%%%%%%%%%%%%%%%%%%%%%%%%%%%%%%%        
 In this approach one considers composite gauge invariant operators in          
 the standard Legendre transformation framework.                                
 The obvious advantage of this approach is that the potential can               
 be evaluated perturbatively and, since it is gauge invariant, it can           
 be directly compared to lattice simulations.                                   
 It offers therefore a conceptually clean and directly accessible               
 tool of confronting results obtained in perturbation theory                    
 with numerical data.                                                           
 In this paper we want to report about our first experiences with the           
 gauge invariant potential.                                                     
 We will use it mainly for the determination of the transition points.          
 We postpone a discussion of its renormalization                                
 and the extraction of physical quantities to a future publication.             
                                                                                
 The starting point for the gauge invariant effective potential                 
 for the length square of the Higgs field is the free energy $F(J)$ in          
 the presence of a constant external source $J$                                 
\be \label{eq3.1}                                                               
e^{-\Omega F(J)} = \int [dU] [d\varphi] e^{-S +J\sum_x \rho_x^2}  \ ,               
\ee                                                                             
 with $S$ the action eq.~(\ref{eq1.1}) and $\Omega$ the lattice volume.        
 From this the effective potential                                              
 is obtained by a Legendre transformation                                       
\be \label{eq3.2}                                                               
V (\bar{\rho}^2) = F (J(\bar{\rho}^2)) - \bar{\rho}^2 J  \ ,                     
\ee                                                                             
where                                                                           
\be \label{eq3.3}                                                               
\bar{\rho}^2 = \frac{\partial}{\partial J} F(J)\; .                             
\ee                                                                             
                                                                                
 The perturbative evaluation of the gauge invariant effective                   
 potential is done in the standard loopwise semiclassical expansion.            
 To proceed we choose the unitary gauge and set the angle of the                
 Higgs field to $\alpha_x = {\bf 1}$.                                                  
 Since the free energy is gauge invariant this has no effect on the             
 effective potential.                                                           
 The action becomes                                                             
\be \label{eq3.4}                                                               
S = S_g                                                                         
+ \sum_x \left\{ \rho_x^2 +                                                     
\lambda \left[ \rho_x^2 - 1 \right]^2 -                                         
\kappa\sum_{\mu=1}^4                                                            
\rho_{x+\hat{\mu}}\rho_x {\rm Tr\,}U_{x\mu}                                     
\right\}  \ ,                                                                       
\ee                                                                             
 with $S_g$ the plaquette action for the gauge field defined in                 
 (\ref{eq1.1}).                                                                 
                                                                                
 A stationary point is obtained for                                             
\be \label{eq3.5}                                                               
J > 1 - 8\kappa -2\lambda                                                       
\ee                                                                             
 and $U_{x,\mu}$ the unit matrix as                                             
\be \label{eq3.6}                                                               
\bar{\rho}^2 = \frac{J+8\kappa -1}{2\lambda} +1 \;.                             
\ee                                                                             
 The last equation is easily inverted for $J(\bar{\rho}^2)$ and the             
 effective potential to tree level is                                           
\be \label{eq3.7}                                                               
V_{tree}(\bar{\rho}^2) = (1-8\kappa) \bar{\rho}^2                               
+\lambda (\bar{\rho}^2-1)^2.                                                    
\ee                                                                             
 To get the one-loop effective potential we have to consider                    
 fluctuations around the stationary point (\ref{eq3.6}).                        
 The fluctuations to one-loop consist of a gauge part and a Higgs part          
 and at this level no mixing appears.                                           
 One obtains                                                                    
\be \label{eq3.8}                                                               
V_{1-loop} = V_{tree}                                                           
+ \int_{-\pi}^{\pi}\frac{d^4 k}{(2\pi )^4}\left\{                               
  \frac{9}{2} \ln (\hat{k}^2 + m_g^2)                                           
+ \frac{1}{2} \ln (\hat{k}^2 + m_\phi^2 ) \right\} \ ,                          
\ee                                                                             
 where the masses are related to the parameters in the lattice action           
 (\ref{eq3.4}) and $\bar{\rho}^2$ by                                            
\be  \label{eq3.9}
m_g^2   =  \frac{1}{2}\kappa g^2 \bar{\rho}^2  \  ,  \hspace{3em}
m^2_\phi  =  \frac{4}{\kappa}\lambda \bar{\rho}^2                             
\ee
 and the momenta in the lattice integrals are                                   
 $\hat{k}^2 = \sum_\mu[2-2\cos (k_\mu )]$.                                      
                                                                                
 The solution of the effective potential given above corresponds                
 to the broken phase of the SU(2)-Higgs model.                                  
 In \cite{BUFOHE} it was emphazised that there exists another                   
 stationary point which belongs to the symmetric phase                          
 of the model and which is given by $\bar{\rho} = 0$.                           
 In this case the tree level potential is trivially zero and we                 
 have to start with the one-loop formula for the free energy                    
\be \label{eq3.10}                                                              
F(J)=\frac{1}{2}\int_{-\pi}^{\pi}\frac{d^4 k}{(2\pi )^4}                        
            \ln (\hat{k}^2 + m_0^2) \ ,                                             
\ee                                                                             
 with $m_0^2 = (1-8\kappa-2\lambda)/\kappa + J$.                                
 To obtain the effective potential one has to solve                             
 eq.~(\ref{eq3.3}) for $J(\bar{\rho}^2)$.                                      
 In \cite{BUFOHE} the solution has been given for the                           
 three dimensional Higgs model in a closed form.                                
 A description of the effective potential in the symmetric                      
 phase has been found with a quite characteristic asymmetric shape.             
                                                                                
 In our case we have to work with lattice integrals or finite                   
 lattice sums.                                                                  
 Then the solution can no longer be given in a closed form.                     
 However, one can perform the Legendre transformation and solve                 
 eq.~(\ref{eq3.3}) numerically.                                                
 The result of this procedure for the points where our simulations              
 are performed confirm the general shape of the                                 
 potential in the symmetric phase and are qualitatively in agreement            
 with the potential extracted from the distributions of $\rho^2$ values         
 from the simulations.                                                          
 However, we do not have a quantitative understanding                           
 of the symmetric phase yet.                                                    
 We hope to come back to this question in a future publication.                 
                                                                                
 A final remark concerns the lattice simulations where the gauge                
 invariant effective potential is obtained from a distribution of the           
 operator under consideration.                                                  
 The potential computed in this way is the so-called                            
 constraint effective potential \cite{ORAWIYO}.                                 
 In the infinite volume limit this potential coincides with the one             
 defined by means of the Legendre transformation above.                         
 In perturbation theory both approaches differ in the treatment of the          
 zero modes.                                                                    
 For the one-loop result (\ref{eq3.8}) this amounts to leaving out the          
 $k=0$ mode in the finite lattice sums for the constraint effective             
 potential.                                                                     
                                                                                
%%%%%%%%%%%%%%%%%%%%%%%%%%%%%%%%%%%%%%%%%%%%%%%%%%%%%%%%%%%%%%%%%%%%%%%%        
                                                                                
\section{Phase transition points}              \label{sec4}                     
 A numerical simulation of the SU(2)-Higgs model should start by first          
 determining the phase transition points.                                       
 Physically the transition is triggered by a temperature change.                
 Keeping all other parameters fixed, this could only be achieved                
 on the lattice by asymmetric couplings, which is possible but                  
 cumbersome.                                                                    
 It will become clear later (see sections \ref{sec5} and \ref{sec6})            
 that in the parameter range we are interested in $\beta$ and                   
 $\lambda$ are fixing, to a good approximation, the renormalized                
 parameters.                                                                    
 A change of $\kappa$ is reflected mainly in a change of the lattice            
 spacing $a$.                                                                   
 Therefore if one crosses the transition at fixed $\beta,\;\lambda$ by          
 changing $\kappa$, the essential change is in the physical temperature         
 $T = 1/(aL_t)$.                                                                
 (The physical volume is assumed to be large enough such that its change        
 with $a^3$ is not important.)                                                  
 Thus we are looking for the phase transition in the hopping parameter          
 at $\kappa=\kappa_c$, for fixed $\beta,\;\lambda$.                             
                                                                                
%%%%%%%%%%%%%%%%%%%%%%%%%%%%%%%%%%%%%%%%%%%%%%%%%%%%%%%%%%%%%%%%%%%%%%%%        
\subsection{One-loop effective potential and                                    
            transition points}                       \label{sec4.1}             
%%%%%%%%%%%%%%%%%%%%%%%%%%%%%%%%%%%%%%%%%%%%%%%%%%%%%%%%%%%%%%%%%%%%%%%%        
\begin{figure}                                                                  
\vspace{9.0cm}                                                                  
\includegraphics{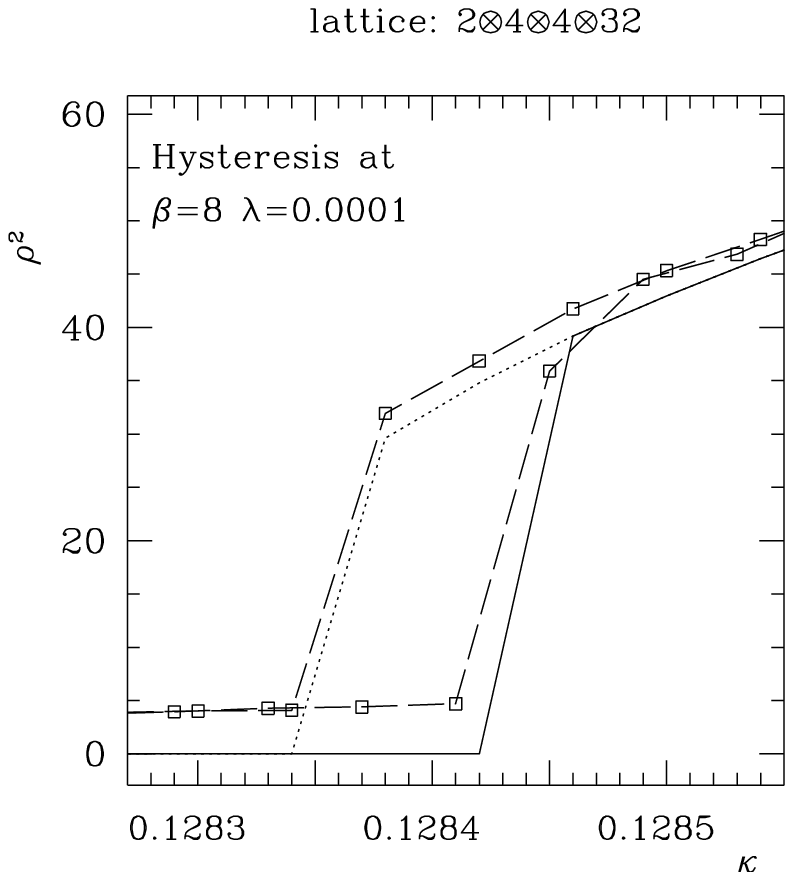}                                               
\begin{center}                                                                  
\parbox{15cm}{\caption{ \label{fig4.1}                                          
 Thermal cycle exhibiting a hysteresis in $\rho^2$.                             
 The solid line indicates the values of the absolut minima from the             
 gauge invariant effective potential, the                                       
 short dashed line the ones of the false minima.                                
 The long dashed line only connects the data points to guide the eye.           
}}                                                                              
\end{center}                                                                    
\end{figure}                                                                    
%%%%%%%%%%%%%%%%%%%%%%%%%%%%%%%%%%%%%%%%%%%%%%%%%%%%%%%%%%%%%%%%%%%%%%%%        
 We found that for searching the transition point the gauge invariant           
 effective potential can be very helpful.                                       
 It can serve as a tool to provide quite accurate information about             
 $\kappa_c$ which helps to select the $\kappa$-values where simulations         
 are then performed.                                                            
 We define the transition point $\kappa_c$ as the $\kappa$-value where          
 the symmetric and broken minima of the gauge invariant effective               
 potential are degenerate.                                                      
 For the computation we used the one-loop formula eq.~(\ref{eq3.8})            
 for the broken phase and the trivial, $\rho^2 =0$, minimum for the             
 symmetric phase.                                                               
 Although this is certainly not the exact value for the symmetric               
 minimum, we will see in the following that for small $\lambda$ the             
 transition $\kappa's$ are in very good agreement with numerical data.          
                                                                                
 For the computation of the gauge invariant effective potential                 
 on a finite lattice of size $L_x \cdot L_y \cdot L_z \cdot L_t$ the            
 integrals in (\ref{eq3.8}) have been replaced by the corresponding             
 lattice sums.                                                                  
 Following the experience in QCD \cite{LEMA}, we also used the                  
 mean field improved gauge coupling $g\rightarrow g/\sqrt{U_{pl}}$,             
 with $U_{pl}$ the measured plaquette value.                                    
 This choice for the gauge coupling  appeared to be very useful to              
 achieve better agreement with simulation results.                              
                                                                                
 At the values of the Higgs mass where our simulations are performed            
 the electroweak phasetransition is certainly of first order.                   
 The effective potential is such that besides the                
 absolute minimum a secondary minimum exists.                                   
 In the simulations this phenomenon leads to                                    
 hysteresis effects when the system gets stuck in the wrong minimum.            
 Therefore hysteresis effects in thermal cycles are often taken                 
 as an indication for a first order phase transition.                           
 The effective potential allows to compute the values of $\rho^2$               
 also in the false vacuum and  should hence reproduce the hysteresis.           
 In fig.~\ref{fig4.1} we show a thermal cycle at $\beta  =8$,                  
 $\lambda=0.0001$ on a $2\cdot 4^2\cdot 32$ lattice.                            
 The data points, connected by a long dashed line to guide the eye,             
 show a clear hysteresis.                                                       
 The solid line represents the values of $\rho^2$ in the absolut                
 minimum obtained from the gauge invariant effective potential.                 
 The short dashed line indicate the values of $\rho^2$ in the second            
 minimum.                                                                       
 The figure demonstrates that the numerical data are very well                  
 described by the one-loop gauge invariant effective potential.                 
 The agreement gets worse for the high point at $\lambda =0.0005$.              
 Here the transition points from the perturbatively                             
 evaluated potential and the simulations show some discrepancy, see             
 table \ref{2kaptab}.                                                           
                                                                                
 In section \ref{sec7} we will compute the latent heat.                         
 For this we need transition $\kappa's$ for $L_t = 2,...,5$.                    
 For the higher $L_t=4,5$-values numerical simulations are very                 
 demanding as one would have to scale the other extensions of the               
 lattice accordingly.                                                           
 Therefore we will resort there to the values of $\kappa_c$ as                 
 obtained from the effective potential.                                         
 For this purpose we performed a finite size scaling analysis of                
 $\kappa_c^L$ on various size lattices                                          
\be \label{eq4.1}                                                               
\kappa_c^L = a V^{-\nu} + \kappa_c^\infty   \ ,                                 
\ee                                                                             
 where $V = L_x \cdot L_y \cdot L_z$ and $L_t$ is kept fixed.                   
 We computed $\kappa_c^L$ for various $V$ from the gauge invariant              
 effective potential and fitted it to (\ref{eq4.1}).                            
 In all the fits performed, we found a value of $\nu = 1.00(2)$.                
 This again confirms the first order nature of the electroweak                  
 phase transition for Higgs masses below 50 GeV.                                
 The obtained results for $\kappa_c^\infty$ for various $L_t$, $\beta$          
 and $\lambda$ are plotted in fig.~\ref{fig6.3} where we discuss               
 the lines of constant physics.                                                 
 The numerical values in our four basic points are:                             
 $l2:\;\kappa_c^\infty=0.128290(1)$;                                            
 $l3:\;\kappa_c^\infty=0.128082(1)$;                                            
 $h2:\;\kappa_c^\infty=0.128625(1)$;                                            
 $h3:\;\kappa_c^\infty=0.128273(1)$.                                            
                                                                                
%%%%%%%%%%%%%%%%%%%%%%%%%%%%%%%%%%%%%%%%%%%%%%%%%%%%%%%%%%%%%%%%%%%%%%%%        
\subsection{Two-coupling method and transition points}\label{sec4.2}            
 Provided hysteresis effects in thermal cycles are seen at a first              
 order phase transition, the two-coupling method is useful for a                
 precise determination of the position of the phase transition point.           
                                                                                
 Let us consider the largest extension of the lattice to be the                 
 $z$-direction.                                                                 
 In this direction the lattice is divided into two halves.                      
 The idea is to choose different coupling parameters in both halves,            
 to enforce one part to stay in the symmetric phase and the other               
 one in the Higgs phase.                                                        
 We assume that the $z$-direction is long enough for accomodating a             
 pair of interfaces between the phases.                                         
 (If the $z$-direction is not too long, the appearance of more                  
 interface pairs has a negligible probability.)                                 
 By warming up the configuration far enough in both directions from the         
 transition point one can achieve that at the beginning of the                  
 simulation at the envisaged pair of parameters both phases and the             
 interface pair are present.                                                    
 In this configuration the system can sensitively react to free                 
 energy differences by shifts of the interface positions in the                 
 $z$-direction.                                                                 
 If the configuration stays for many autocorrelation times in the               
 mixed state the free energies of the two phases at the chosen                  
 parameters are such that this situation is stable against                      
 transitions to a unique phase.                                                 
 Thus the parameter sets of the two phases give a lower and an upper            
 bound on the transition parameters.                                            
 In fact, due to the additional free energy associated with the                 
 interfaces, the phase transition from a two-phase situation to a               
 unique phase occurs somewhat earlier than the equality of the free             
 energies of the two phases.                                                    
 This parameter shift goes, however, exponentially to zero for                  
 increasing lattice volumes.                                                    
                                                                                
 Of course, the two-coupling method for the determination of the                
 transition point works only if one is able to tell one phase                   
 from the other.                                                                
 A possible way is to perform hysteresis runs on lattices with the              
 same extensions as half of the lattice for the two-coupling method.            
 The hysteresis plots are used for the distinction of the two phases            
 at a given $\kappa$.                                                           
                                                                                
 As stated in the introduction of this section, we are interested in            
 the position of the phase transition in the hopping parameter at               
 $\kappa=\kappa_c$ for fixed $\beta$ and $\lambda$.                             
 Thus the only parameter chosen differently in both parts of the                
 lattice is the hopping parameter.                                              
 At each pair of $\kappa$ the system was observed for at least 10               
 autocorrelation times.                                                         
 $\kappa_c$ was defined as the mean value of the best lower and upper           
 bounds.                                                                        
 The best estimates for the transition point $\kappa_c$ obtained with           
 this method are given in table~\ref{2kaptab}.                                  
%%%%%%%%%%%%%%%%%%%%%%%%%%%%%%%%%%%%%%%%%%%%%%%%%%%%%%%%%%%%%%%%%%%%%%%         
\begin{table}                                                                   
\begin{center}                                                                  
\parbox{15cm}{\caption{ \label{2kaptab}                                         
 Hopping parameter $\kappa_c$ at the transition point, obtained with            
 the two coupling method.                                                       
 As lattice size, the total size of both parts is given.                        
 For comparison, the estimates obtained on lattices with half the               
 size from the one-loop invariant effective potential are also given.           
}}                                                                              
\end{center}                                                                    
\begin{center}                                                                  
\begin{tabular}%                                                                
{|c@{$\,\cdot\,$}c@{$\,\cdot\,$}c@{$\,\cdot\,$}c|l|l||l|l|}                     
\hline                                                                          
\multicolumn{4}{|c|}{Lattice}&\multicolumn{1}{c|}{$\beta$}                      
&\multicolumn{1}{c||}{$\lambda$}&\multicolumn{1}{c|}{$\kappa_c$}                
&\multicolumn{1}{c|}{$\kappa_c^{1-loop}$}\\                                     
\hline\hline                                                                    
2  &   4  &   4  &  128 &   8.0   &  0.0001   &  0.12836(3) &                   
0.12840(5)       \\ \hline                                                      
2  &  16  &  16  &  128 &   8.0   &  0.0001   &  0.12830(5) &                   
0.12828(2)       \\ \hline                                                      
3  &   6  &   6  &  192 &   8.15  &  0.00011  &  0.12813(2) &                   
0.12811(2)       \\ \hline                                                      
3  &  12  &  12  &  192 &   8.15  &  0.00011  &  0.12811(1) &                   
0.12811(1)       \\ \hline                                                      
2  &  32  &  32  &  256 &   8.0   &  0.0005   &  0.12887(1) &                   
0.12862(1)       \\ \hline                                                      
3  &  48  &  48  &  384 &   8.15  &  0.00051  &  0.12852(2) &                   
0.12826(1)       \\ \hline                                                      
\end{tabular}                                                                   
\end{center}                                                                    
\end{table}                                                                     
%%%%%%%%%%%%%%%%%%%%%%%%%%%%%%%%%%%%%%%%%%%%%%%%%%%%%%%%%%%%%%%%%%%%%%%         
                                                                                
%%%%%%%%%%%%%%%%%%%%%%%%%%%%%%%%%%%%%%%%%%%%%%%%%%%%%%%%%%%%%%%%%%%%%%%%        
\subsection{Multicanonical method and transition points}\label{sec4.3}          
 The most precise way to determine the transition point has                     
 been provided by a combination of the multicanonical method and                
 $\kappa$-reweighting \cite{FERRENBERG}.                                        
 In this method first an order parameter distribution is generated              
 through a Monte Carlo simulation.                                              
 This distribution is then extrapolated to nearby $\kappa$'s                    
 to find the transition point.                                                  
 The action with the additional logarithmic term ($S_{log}$ in                  
 eq.~(\ref{slog})) and the link-variable ($L_\varphi$ in                       
 eq.~(\ref{eq5.2})) have been considered as order parameters.                  
 The transition point is  determined by the equal height                        
 signal: $\kappa_c$ is the hopping parameter value for which                    
 the heights of the two peaks in the distribution are equal.                    
 The systematic uncertainty of $\kappa_c$ can be estimated by                   
 comparing its values obtained by equal height signal and equal area            
 signal for different order parameters.                                         
                                                                                
 The multicanonical simulation has been performed at $\kappa=\kappa_v$,         
 close to the real transition point                                             
 ($\vert \kappa_v - \kappa_c \vert = {\cal O}(10 ^{-5})$).                      
 The approximate transition point $\kappa_v$ was determined by the              
 one-loop invariant effective potential.                                        
 The distribution of an order parameter at a nearby $\kappa'$, with             
 all other couplings fixed, can be obtained by attaching a weight               
\be \label{reweight}                                                            
w(L_\varphi)=e^{8\Omega L_\varphi(\kappa'-\kappa_v)}                            
\ee                                                                             
 to the different configurations.                                               
                                                                                
\begin{figure}                                                                  
\vspace{9.0cm}                                                                  
\includegraphics{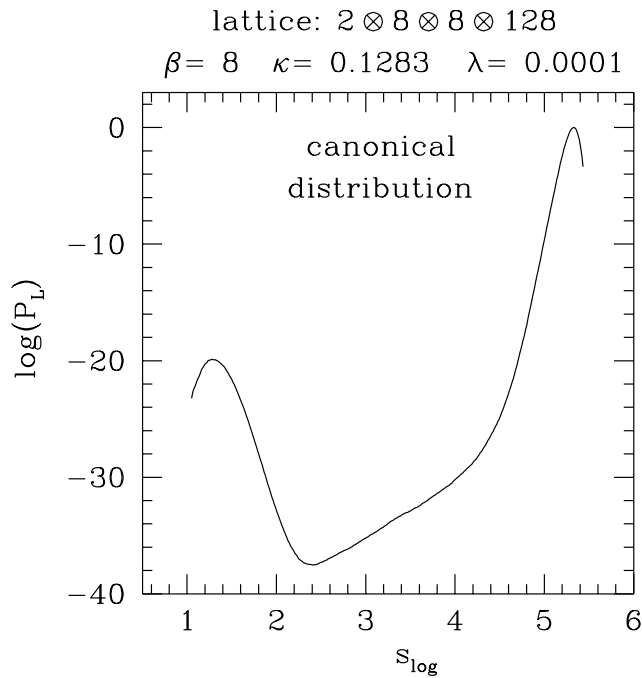}                                                  
\begin{center}                                                                  
\parbox{15cm}{\caption{ \label{fig4.2}                                          
 The distribution of the action density obtained from a                         
 constrained-multicanonical simulation on $2\cdot 8^2  \cdot 128$               
 lattice.
 Note that the left hand peak corresponds to the Higgs phase
 and the right hand one to the symmetric phase.
}}                                                                              
\end{center}                                                                    
\end{figure}

 The action density distribution at the {\it low} point for                     
 $\kappa_v=0.1283$  on a lattice with $\Omega=2\cdot 8^2  \cdot 128$            
 sites is shown in fig.~\ref{fig4.2}.                                          
%%%%%%%%%%%%%%%%%%%%%%%%%%%%%%%%%%%%%%%%%%%%%%%%%%%%%%%%%%%%%%%%%%%%%%%%        
%%%%%%%%%%%%%%%%%%%%%%%%%%%%%%%%%%%%%%%%%%%%%%%%%%%%%%%%%%%%%%%%%%%%%%%%        
%%%%%%%%%%%%%%%%%%%%%%%%%%%%%%%%%%%%%%%%%%%%%%%%%%%%%%%%%%%%%%%%%%%%%%%%        
\begin{figure}                                                                  
\vspace{9.0cm}                                                                  
\includegraphics{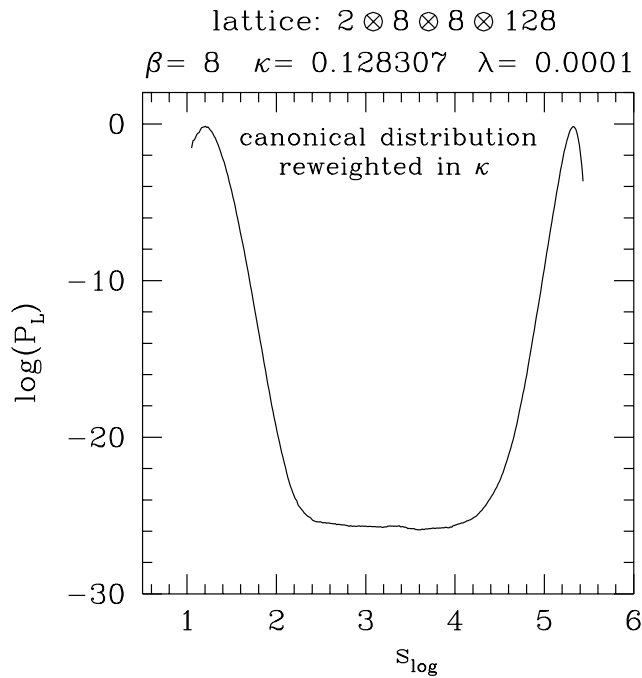}                                              
\begin{center}                                                                  
\parbox{15cm}{\caption{ \label{fig4.3}                                          
 The reweighted distribution of the action density obtained from                
 the data of the previous figure at $\kappa=\kappa_c=0.128307$.                 
}}                                                                              
\end{center}                                                                    
\end{figure}                                                                    
%%%%%%%%%%%%%%%%%%%%%%%%%%%%%%%%%%%%%%%%%%%%%%%%%%%%%%%%%%%%%%%%%%%%%%%%        
 The distribution in neighbouring points is obtained by                         
 eq.~(\ref{reweight}).                                                         
 The two peaks are of equal height at $\kappa_c=0.128307$                       
 (fig.~\ref{fig4.3}).                                                          
 As it can be seen, not only the heights are equal                              
 but also the widths of the two peaks are quite similar, thus the               
 equal height condition for $\kappa_c$ is roughly equivalent to the             
 equal area condition.                                                          
 At the same time the flat regime between the peaks is almost constant.         
 This means that in a multicanonical simulation the two phases can mix          
 with each other with an arbitrary mixing ratio.                                
 The supression is due to the interfaces between the phases.                    
 Taking $L_\varphi$ as an order parameter, the transition point                 
 corresponding to the equal height signal is given by a slightly higher         
 $\kappa$=0.128308.                                                             
 This leads to a transition point at $\kappa_c=0.128307(2+1)$,                  
 where the first number in brackets denotes the systematic the second           
 one the statistical error estimate.                                            
 Similarly for $\Omega=2\cdot 4^2  \cdot 128$ and for                           
 $\Omega=2\cdot 4^2  \cdot 64$ the transition points are at                     
 $\kappa_c=0.128366(3+1)$ and $\kappa_c=0.128367(3+1)$, respectively.

%%%%%%%%%%%%%%%%%%%%%%%%%%%%%%%%%%%%%%%%%%%%%%%%%%%%%%%%%%%%%%%%%%%%%%%%        
                                                                                
\section{Masses and correlation lengths}       \label{sec5}                     
 Important characteristic features of any statistical physical                  
 system are the correlation lengths.                                            
 At zero temperature their inverses give the masses of the low                  
 lying particles.                                                               
 Particularly interesting are also the (inverse) correlation lengths on         
 the two sides of the phase transition.                                         
                                                                                
%%%%%%%%%%%%%%%%%%%%%%%%%%%%%%%%%%%%%%%%%%%%%%%%%%%%%%%%%%%%%%%%%%%%%%%%        
\subsection{Zero temperature masses}                 \label{sec5.1}             
%%%%%%%%%%%%%%%%%%%%%%%%%%%%%%%%%%%%%%%%%%%%%%%%%%%%%%%%%%%%%%%%%%%%%%%%        
\begin{figure}                                                                  
\vspace{9.0cm}                                                                  
\includegraphics{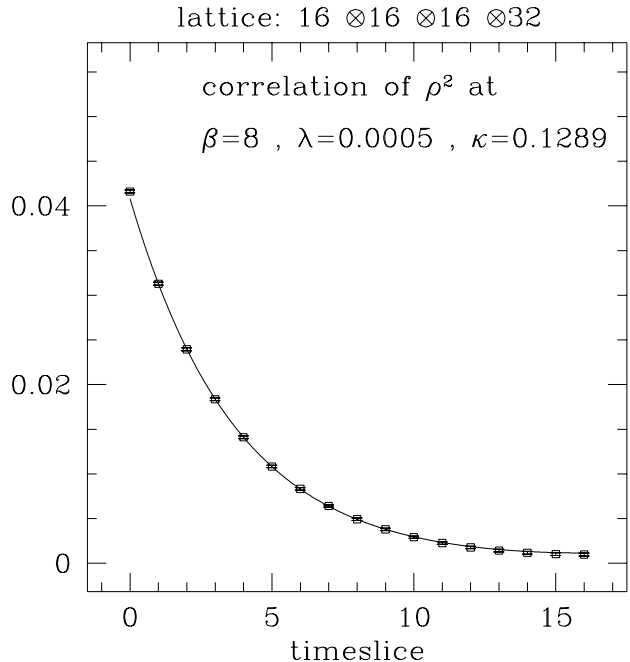}                                         
\begin{center}                                                                  
\parbox{15cm}{\caption{ \label{fig5.1}                                          
 Correlation function in the Higgs boson channel in point $h2[16/90]$.          
 The curve shown is the best fit for timeslices between 1 and 11                
 with $M_H=0.2663$ and a constant factor $f_H=4.081\; 10^{-2}$.                 
 The $\chi$-square of this fit is $\chi^2=0.54$.                                
 The statistical error of the correlation function at distance 1 is             
 about 0.5\%, at distance 11 about 6\%.                                         
}}                                                                              
\end{center}                                                                    
\end{figure}                                                                    
%%%%%%%%%%%%%%%%%%%%%%%%%%%%%%%%%%%%%%%%%%%%%%%%%%%%%%%%%%%%%%%%%%%%%%%%        
%%%%%%%%%%%%%%%%%%%%%%%%%%%%%%%%%%%%%%%%%%%%%%%%%%%%%%%%%%%%%%%%%%%%%%%%        
\begin{table}                                                                   
\begin{center}                                                                  
\parbox{15cm}{\caption{ \label{tab5.1}                                          
 The parameter values of numerical simulations for determining                  
 zero temperature masses and Wilson loops.                                      
}}                                                                              
\end{center}                                                                    
\begin{center}                                                                  
\begin{tabular}{|c|c|c|c|c|c|}                                                  
\hline                                                                          
index & lattice & $\beta$ & $\lambda$ & $\kappa$ & sweeps \\                    
\hline\hline                                                                    
$l2$     &  $12^3 \cdot 24$  &                                                  
8.00  &  0.00010  &  0.12830  &  150000  \\                                     
\hline                                                                          
$l3$     &  $18^3 \cdot 36$  &                                                  
8.15  &  0.00011  &  0.12810  &  125000  \\                                     
\hline                                                                          
$h2[12]$  &  $12^3 \cdot 24$  &                                                 
8.00  &  0.00050  &  0.12890  &  160000  \\                                     
\hline                                                                          
$h2[16/85]$  &  $16^3 \cdot 32$  &                                              
8.00  &  0.00050  &  0.12885  &  60000   \\                                     
\hline                                                                          
$h2[16/90]$  &  $16^3 \cdot 32$  &                                              
8.00  &  0.00050  &  0.12890  &  160000  \\                                     
\hline                                                                          
$h3$     &  $18^3 \cdot 36$  &                                                  
8.15  &  0.00051  &  0.12852  &   50000  \\                                     
\hline                                                                          
\end{tabular}                                                                   
\end{center}                                                                    
\end{table}                                                                     
%%%%%%%%%%%%%%%%%%%%%%%%%%%%%%%%%%%%%%%%%%%%%%%%%%%%%%%%%%%%%%%%%%%%%%%%        
%%%%%%%%%%%%%%%%%%%%%%%%%%%%%%%%%%%%%%%%%%%%%%%%%%%%%%%%%%%%%%%%%%%%%%%%        
\begin{figure}                                                                  
\vspace{9.0cm}                                                                  
\includegraphics{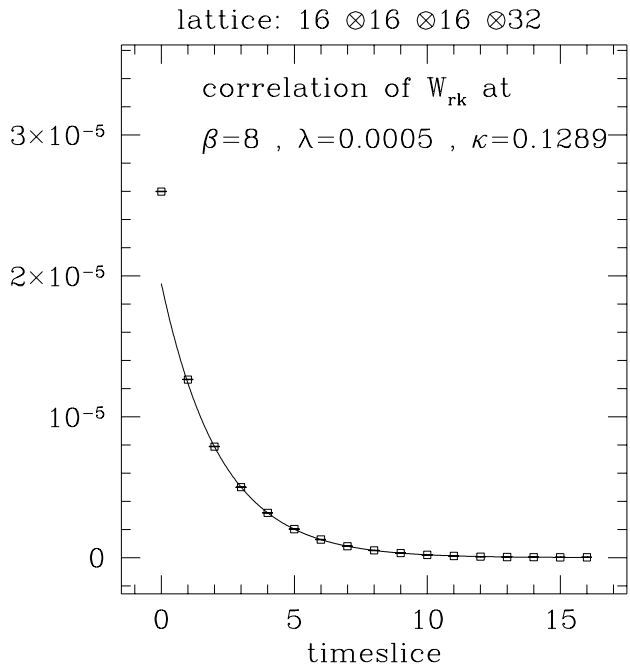}                                           
\begin{center}                                                                  
\parbox{15cm}{\caption{ \label{fig5.2}                                          
 Correlation function in the W-boson channel in point $h2[16/90]$.              
 The curve shown is the best fit for timeslices between 3 and 15                
 with $M_W=0.4522$ and a constant factor $f_W=1.946\; 10^{-5}$.                 
 The $\chi$-square of this fit is $\chi^2=1.30$.                                
 The statistical error of the correlation function at distance 3 is             
 about 0.2\%, at distance 15 about 26\%.                                        
}}                                                                              
\end{center}                                                                    
\end{figure}                                                                    
%%%%%%%%%%%%%%%%%%%%%%%%%%%%%%%%%%%%%%%%%%%%%%%%%%%%%%%%%%%%%%%%%%%%%%%%        
%%%%%%%%%%%%%%%%%%%%%%%%%%%%%%%%%%%%%%%%%%%%%%%%%%%%%%%%%%%%%%%%%%%%%%%%        
\begin{table}                                                                   
\begin{center}                                                                  
\parbox{15cm}{\caption{ \label{tab5.2}                                          
 The W-boson mass $M_W$ and Higgs boson mass $M_H$ in the                       
 points defined by the previous table.                                          
 Their ratio is $R_{HW} \equiv M_H/M_W$ which leads to the                      
 Higgs boson mass in physical units given in the last column.                   
}}                                                                              
\end{center}                                                                    
\begin{center}                                                                  
\begin{tabular}{|c||l|l|l|l|c|}                                                 
\hline                                                                          
index & \multicolumn{1}{c|}{$M_W$}     & \multicolumn{1}{c|}{$M_H$}             
      & \multicolumn{1}{c|}{$R_{HW}$}  & \multicolumn{1}{c|}{$T_c/M_W$}         
      & \multicolumn{1}{c|}{$M_H(GeV)$}        \\                               
\hline\hline                                                                    
$l2$     &                                                                      
1.059(24)   &   0.236(7)  &  0.222(12)  &  0.472(11)  &  18  \\                 
\hline                                                                          
$l3$     &                                                                      
0.718(3)    &  0.144(4)   &  0.201(5)   &  0.464(2)   &  16  \\                 
\hline                                                                          
$h2[12]$  &                                                                     
0.427(8)    &   0.262(9)  &  0.614(32)  &  1.171(22)  &  49  \\                 
\hline                                                                          
$h2[16/85]$  &                                                                  
0.427(5)     &  0.253(8)  &  0.593(19)  &  1.171(19)  &  47  \\                 
\hline                                                                          
$h2[16/90]$  &                                                                  
0.453(3)    &   0.266(5)  &  0.587(12)  &  1.104(7)   &  47  \\                 
\hline                                                                          
$h3$     &                                                                      
0.289(4)    &  0.175(7)   &  0.606(24)  &  1.153(16)  &  48  \\                 
\hline                                                                          
\end{tabular}                                                                   
\end{center}                                                                    
\end{table}                                                                     
%%%%%%%%%%%%%%%%%%%%%%%%%%%%%%%%%%%%%%%%%%%%%%%%%%%%%%%%%%%%%%%%%%%%%%%%        
 The physical Higgs mass $M_H$ can be extracted from correlators                
 of quantities as the site variable                                             
\be \label{eq5.1}                                                               
R_x \equiv \half{\rm Tr\,}(\varphi_x^+\varphi_x)                                
\equiv \rho_x^2 \ ,                                                             
\ee                                                                             
 or, using $\varphi_x \equiv \rho_x \alpha_x$, the link variables               
\be \label{eq5.2}                                                               
L_{\alpha,x\mu} \equiv                                                          
\half {\rm Tr\,}(\alpha^+_{x+\hat{\mu}}U_{x\mu}\alpha_x) \ ,                    
\hspace{3em}                                                                    
L_{\varphi,x\mu} \equiv                                                         
\half {\rm Tr\,}(\varphi^+_{x+\hat{\mu}}U_{x\mu}\varphi_x) \ .                  
\ee                                                                             
 The W-boson mass $M_W$ can be obtained similarly from the composite            
 link fields ($r,k=1,2,3$)                                                      
\be \label{eq5.3}                                                               
W_{xrk} \equiv                                                                  
\half {\rm Tr\,}(\tau_r \alpha^+_{x+\hat{k}}U_{xk}\alpha_x) \ .                 
\ee                                                                             
 Due to lattice symmetry only the diagonal correlators in                       
 $(r,k)$ are non-zero and they can be averaged before extracting                
 masses.                                                                        
                                                                                
 Numerical simulations to determine the zero temperature masses                 
 were performed on $12^3 \cdot 24$ and $16^3 \cdot 32$ lattices                 
 at the phase transition points of $L_t=2$ lattices.                            
 The former were scaled up to $18^3 \cdot 36$ at the corresponding              
 $L_t=3$ points.                                                                
 Since the gauge-Higgs system is in the Higgs phase at low                      
 temperatures and in the symmetric phase at high temperatures,                  
 the $T=0$ points at the transition points are always in the                    
 Higgs phase.                                                                   
 We did neither attempt to investigate the particle spectrum at                 
 $T=0$ in the symmetric phase nor did we determine the                          
 $\kappa$-dependence of the masses in the Higgs phase in a wider range          
 of $\kappa$.                                                                   
 These questions could be addressed in future studies.                          
 The only information on $\kappa$-dependence is obtained from two               
 points on $16^3 \cdot 32$ lattices at nearby $\kappa$-values.                  
 The collection of parameters and lattices where zero temperature               
 masses were determined is contained in table \ref{tab5.1}.                     
 (In the same simulations the Wilson loops were also determined:                
 see section \ref{sec6}.)                                                       
                                                                                
 Masses were extracted from the connected correlators by least square           
 fits by a single $\cosh$ or $\cosh$ + $\rm constant$.                              
 Constant contributions are possible in the Higgs channels but were             
 most of the time negligible, because they are of the order                     
 $\exp(-M_H L_t)$ and our lattices usually have large enough time               
 extension $L_t$.                                                               
 Typical examples in the $h2[16/90]$ point on $16^3 \cdot 32$ are shown         
 by figures \ref{fig5.1} and \ref{fig5.2}.                                      
                                                                                
 Statistical errors on masses were always estimated by subdividing              
 the data sample into subsamples.                                               
 Performing the fits in subsamples gives estimates of standard                  
 deviations of fit parameters.                                                  
 This is particularly straightforward on the Quadrics Q16,                      
 because these lattices can be simulated on 8 nodes                             
 which is repeated 16 times with independent sequences of random                
 numbers.                                                                       
 The 16 parallel sets of statistically independent results can be used          
 for the estimate of fit parameter errors.                                      
 The error estimates in last digits are always given in parentheses.            
 The obtained results are shown in table \ref{tab5.2}.                          
                                                                                
 The masses are in every case well determined.                                  
 The correlators are always dominated by the lowest state.                      
 The statistical errors of the ratios of Higgs- and W-boson masses show         
 that these two masses are not strongly correlated in the data samples.         
 The values of $T_c/M_W$ in table \ref{tab5.2} are obtained                     
 under the assumption that the points are at the finite temperature             
 transition points of $L_t=2$ and $L_t=3$ lattices, respectively.               
 Therefore $T_c=1/2$ and $T_c=1/3$, respectively.                               
                                                                                
 Points $h2[16/85]$ and $h2[16/90]$ are at slightly different $\kappa$          
 values: $\kappa=0.12885$ and $\kappa=0.12890$, respectively, but               
 otherwise identical.                                                           
 Their comparison shows that the ratio of Higgs- to W-boson masses              
 remains almost constant in this $\kappa$-range.                               
 The same holds also for the renormalized gauge coupling (see next              
 section).                                                                      
 The only significant difference is a small change in the overall scale         
 shown by an about 5\% decrease of the masses in lattice units.                 
 This feature is rather helpful because it makes fine tuning of                 
 the scalar hopping parameter $\kappa$ less relevant.                           
                                                                                
 Comparing points $h2[12]$ with $h2[16/90]$ we see that the finite              
 volume effects are not large.                                                  
 The only statistically significant deviation is seen in the                    
 W-boson mass, which is about 5\% smaller on the $12^3 \cdot 24$                
 lattice than on $16^3 \cdot 32$ (for the change of the static                  
 potential see section \ref{sec6}).                                             
 This implies that within our statistical errors the $16^3 \cdot 32$            
 lattice can probably be considered to represent the infinite volume            
 limit.                                                                         
 Nevertheless a systematic study of finite volume dependences in                
 future studies is desirable.                                                   
                                                                                
%%%%%%%%%%%%%%%%%%%%%%%%%%%%%%%%%%%%%%%%%%%%%%%%%%%%%%%%%%%%%%%%%%%%%%%%        
\subsection{Correlation lengths at transition points}\label{sec5.2}             
 An important effect of finite temperatures is the change in                    
 correlation lengths.                                                           
 This is particularly interesting at a first order phase transition,            
 where correlation lengths stay finite and may eventually differ                
 in the two metastable states.                                                  
 The inverse correlation lengths at finite temperatures may also be             
 called ``temperature dependent masses''.                                       
 In order to distinguish them from the real masses we shall denote              
 them by small letters: $m_W$ for the inverse correlation length in              
 the W-boson channel, $m_H$ the one in the Higgs-boson channel.                 
                                                                                
 In our numerical simulations we make use of the fact that on a large           
 enough lattice the metastability at the first order phase transition           
 becomes so strong that in a finite amount of computer time the system          
 stays in the phase it started from.                                            
 At the first sight this seems to lead to mathematically ill-defined            
 averages, because the transition speed of the lattice from one phase           
 to the other does certainly depend on the particular updating                  
 algorithm.                                                                     
 However, as it is well known (and is discussed in detail                       
 in section \ref{sec4}) the two phases can be well distinguished in             
 the distribution of some order parameter which shows two well                  
 separated peaks on a large enough lattice.                                     
 The averages in the two metastable states can be defined to belong             
 to the two peaks in the order parameter distribution.                          
 The advantage of using large lattices is that one can sit right at the         
 phase transition, hence no extrapolation is necessary (see also                
 section \ref{sec7}).                                                           
 This is more important than saving a little on lattice sizes and               
 staying in an unclear situation with an underdeveloped two peak                
 structure, when no clear separation of the two phases is possible.             
                                                                                
 The lattice shapes we consider are typically of the type                       
 $L_t \cdot L_{xy}^2 \cdot L_z$ with $L_t=2,3$ the inverse temperature          
 direction, $L_{xy} \gg L_t$ the ``small'' spatial directions and               
 $L_z \gg L_{xy}$ the ``long'' spatial direction.                               
 The finite volume effects are mainly determined by $L_{xy}$.                   
 The long direction $L_z$ is chosen to force the interfaces between             
 the two phases to be built perpendicular to this direction.                    
 This is useful for the study of the interface tension (see section             
 \ref{sec8}).                                                                   
 The parameters of our numerical simulations are collected in table             
 \ref{tab5.3}.                                                                  
 The results are contained in table \ref{tab5.4}.                               
%%%%%%%%%%%%%%%%%%%%%%%%%%%%%%%%%%%%%%%%%%%%%%%%%%%%%%%%%%%%%%%%%%%%%%%%        
\begin{figure}                                                                  
\vspace{9.0cm}                                                                  
\includegraphics{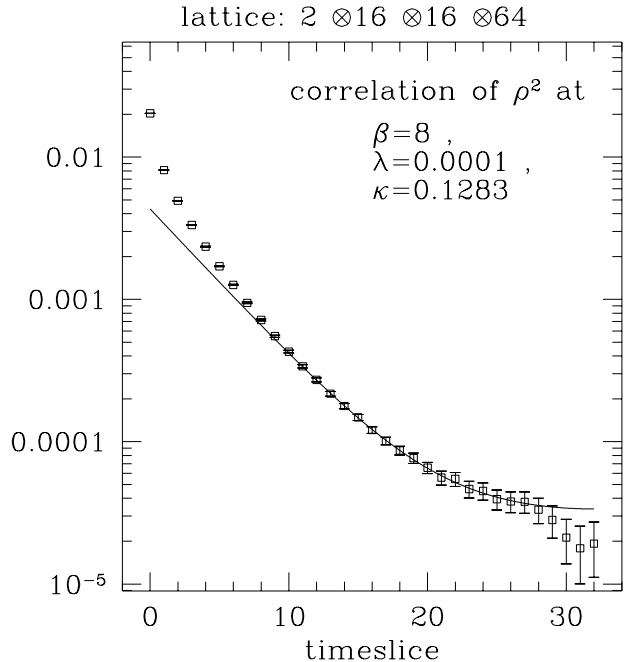}                                        
\begin{center}                                                                  
\parbox{15cm}{\caption{ \label{fig5.3}                                          
 Correlation function in the Higgs boson channel in point $l2[b]$               
 in the symmetric phase.                                                        
 The curve shown is the best fit for timeslices between 12 and 28               
 with $m_H=0.2400$ a constant factor $f_H=4.287\; 10^{-3}$                      
 and an additive constant $c_H=2.974\; 10^{-5}$.                                
 The $\chi$-square of this fit is $\chi^2=0.84$.                                
 The statistical error of the correlation function at distance 12 is            
 about 3\%, at distance 28 about 20\%.                                          
}}                                                                              
\end{center}                                                                    
\end{figure}                                                                    
%%%%%%%%%%%%%%%%%%%%%%%%%%%%%%%%%%%%%%%%%%%%%%%%%%%%%%%%%%%%%%%%%%%%%%%%        
%%%%%%%%%%%%%%%%%%%%%%%%%%%%%%%%%%%%%%%%%%%%%%%%%%%%%%%%%%%%%%%%%%%%%%%%        
\begin{table}                                                                   
\begin{center}                                                                  
\parbox{15cm}{\caption{ \label{tab5.3}                                          
 The parameter values of numerical simulations for determining                  
 inverse correlation lengths at the phase transition.                           
 Indices $[a]$ and $[b]$ refer to simulations started ``above'',                
 respectively, ``below'' the phase transition in $\kappa$-parameter,            
 that is in the Higgs phase, respectively, symmetric phase.                     
}}                                                                              
\end{center}                                                                    
\begin{center}                                                                  
\begin{tabular}{|c|c|c|c|c|c|c|}                                                
\hline                                                                          
index & lattice & $\beta$ & $\lambda$ & $\kappa$ &                              
$a$-sweeps & $b$-sweeps  \\                                                     
\hline\hline                                                                    
$l2[a,b]$ &  $2 \cdot 16^2 \cdot 64$  &                                         
8.00  &  0.00010  &  0.12830  &  320000  &  960000  \\                          
\hline                                                                          
$l3[a,b]$ &  $3 \cdot 24^2 \cdot 96$  &                                         
8.15  &  0.00011  &  0.12810  &  144000  &  80000  \\                           
\hline                                                                          
$h2[a,b]$ &  $2 \cdot 64^2 \cdot 128$  &                                        
8.00  &  0.00050  &  0.12887  &  13000  &  14000  \\                            
\hline                                                                          
$h3[a,b]$ &  $3 \cdot 96^2 \cdot 192$  &                                        
8.15  &  0.00051  &  0.12852  &    5000  &   5000  \\                           
\hline                                                                          
\end{tabular}                                                                   
\end{center}                                                                    
\end{table}                                                                     
%%%%%%%%%%%%%%%%%%%%%%%%%%%%%%%%%%%%%%%%%%%%%%%%%%%%%%%%%%%%%%%%%%%%%%%%        
%%%%%%%%%%%%%%%%%%%%%%%%%%%%%%%%%%%%%%%%%%%%%%%%%%%%%%%%%%%%%%%%%%%%%%%%        
\begin{table}                                                                   
\begin{center}                                                                  
\parbox{15cm}{\caption{ \label{tab5.4}                                          
 The inverse correlation length in the W-channel $m_W$                          
 and in the Higgs channel $m_H$ at the phase transition in the points           
 defined by the previous table.                                                 
 Their ratio is $r_{HW} \equiv m_H/m_W$.                                        
 The physical scale is set by the last column, where the ratio                  
 of $m_W$ to the corresponding zero temperature mass $M_W$ is                   
 given.                                                                         
}}                                                                              
\end{center}                                                                    
\begin{center}                                                                  
\begin{tabular}{|c||l|l|l|l|l|}                                                 
\hline                                                                          
index & \multicolumn{1}{c|}{$m_W$}     & \multicolumn{1}{c|}{$m_H$}             
      & \multicolumn{1}{c|}{$r_{HW}$}  & \multicolumn{1}{c|}{$m_W/T_c$}         
      & \multicolumn{1}{c|}{$m_W/M_W$} \\                                       
\hline\hline                                                                    
$l2[a]$  &                                                                      
0.880(15)   &   0.130(3)  &  0.148(6)   &  1.76(3)    &  0.83(3)    \\          
\hline                                                                          
$l2[b]$  &                                                                      
0.55(4)     &   0.24(3)   &  0.44(9)    &  1.10(8)    &  0.52(5)    \\          
\hline                                                                          
$l3[a]$  &                                                                      
0.569(8)    &  0.0630(27) &  0.111(7)   &  1.71(3)    &  0.80(2)    \\          
\hline                                                                          
$l3[b]$  &                                                                      
0.44(5)     &   0.19(2)   &  0.43(9)    &  0.88(10)   &  0.62(7)    \\          
\hline                                                                          
$h2[a]$  &                                                                      
0.218(6)    &  0.0784(7)  &  0.360(13)  &  0.436(12)  &  0.50(2)    \\          
\hline                                                                          
$h2[b]$  &                                                                      
0.34(6)     &   0.101(12) &  0.30(9)    &  0.68(12)   &  0.80(16)   \\          
\hline                                                                          
$h3[a]$  &                                                                      
0.176(15)   &  0.059(4)   &  0.34(3)    &  0.53(4)    &  0.61(6)    \\          
\hline                                                                          
$h3[b]$  &                                                                      
0.34(7)     & 0.0635(18)  &  0.19(5)    &  1.0(2)     &  1.2(3)     \\          
\hline                                                                          
\end{tabular}                                                                   
\end{center}                                                                    
\end{table}                                                                     
%%%%%%%%%%%%%%%%%%%%%%%%%%%%%%%%%%%%%%%%%%%%%%%%%%%%%%%%%%%%%%%%%%%%%%%%        
%%%%%%%%%%%%%%%%%%%%%%%%%%%%%%%%%%%%%%%%%%%%%%%%%%%%%%%%%%%%%%%%%%%%%%%%        
\begin{figure}                                                                  
\vspace{9.0cm}                                                                  
\includegraphics{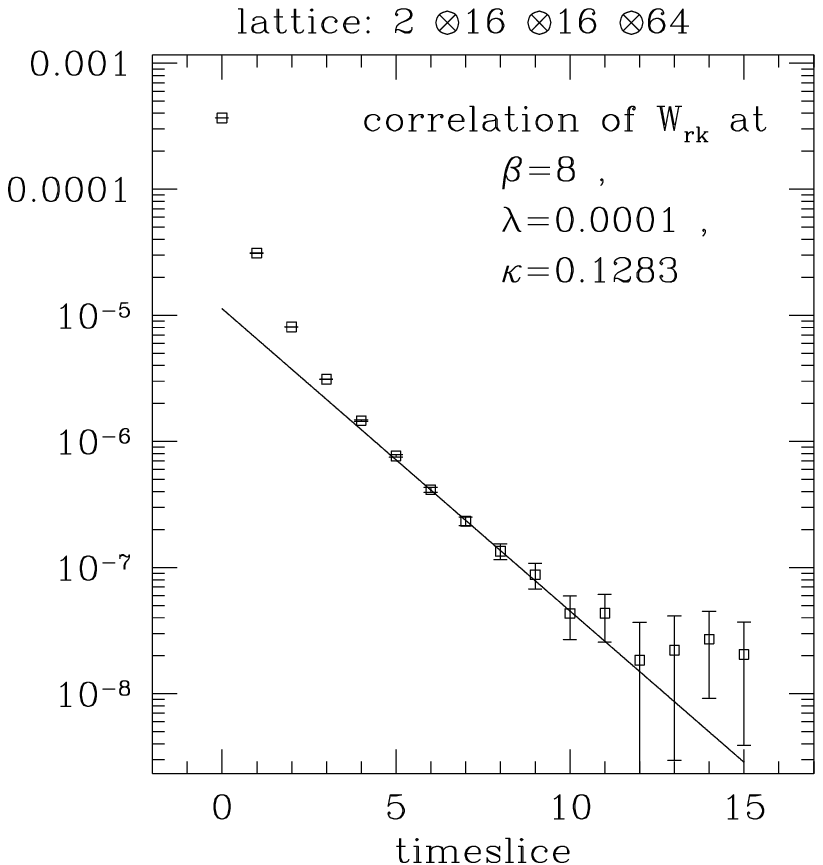}                                          
\begin{center}                                                                  
\parbox{15cm}{\caption{ \label{fig5.4}                                          
 Correlation function in the W-boson channel in point $l2[b]$                   
 in the symmetric phase.                                                        
 The curve shown is the best fit for timeslices between 6 and 10                
 with $m_W=0.5523$ and a constant factor $f_W=1.130\; 10^{-5}$.                 
 The $\chi$-square of this fit is $\chi^2=0.28$.                                
 The statistical error of the correlation function at distance 6 is             
 about 4\%, at distance 10 about 35\%.                                          
}}                                                                              
\end{center}                                                                    
\end{figure}                                                                    
%%%%%%%%%%%%%%%%%%%%%%%%%%%%%%%%%%%%%%%%%%%%%%%%%%%%%%%%%%%%%%%%%%%%%%%%        
                                                                                
 The behaviour of the correlators in the Higgs phase (points with index         
 $a$) is in general similar to the zero temperature points discussed            
 in the previous subsection.                                                    
 There is, however, a substantial difference in the symmetric phase:            
 in general the determination of the inverse correlation lengths is             
 much more difficult.                                                           
 The main reason is that the correlators are not dominated by the               
 contribution of the lowest state.                                              
 There is obviously a dense spectrum contributing in both channels.             
 This feature is particularly pronounced in the W-channel.                      
 For this reason a rather high statistics was taken in the point                
 with index $l2[b]$ (see table \ref{tab5.3}).                                   
 Even in this case the lowest mass in the W-boson channel is still              
 rather unprecise.                                                              
 For an illustration of the errors and fits see figures \ref{fig5.3}            
 and \ref{fig5.4}.                                                              
                                                                                
 There is a clear signal for a discontinuity of $m_W$ and $m_H$                 
 between the two sides of the phase transition, whereby the most                
 significant change is an increase of $m_H$ when going from the                 
 Higgs to the symmetric phase.                                                  
 The values of $m_{W,H}$ are rather small, therefore the lattice                
 volumes may be too small, especially at the $l2,\; l3$ points.                 
 There the infinite volume values may still be somewhat different.              
 From this point of view the situation is much better at the ``high''           
 points, where the volumes had to be chosen much larger due to the              
 smaller value of the interface tension.                                        
                                                                                
%%%%%%%%%%%%%%%%%%%%%%%%%%%%%%%%%%%%%%%%%%%%%%%%%%%%%%%%%%%%%%%%%%%%%%%%        
                                                                                
\section{Renormalized couplings}               \label{sec6}                     
 In the SU(2) Higgs model there are two independent dimensionless               
 renormalized couplings which are conventionally defined at zero                
 temperature.                                                                   
 Corresponding to the bare gauge coupling $g$ and bare quartic coupling         
 $\lambda$ one can introduce the renormalized couplings $g_R$,                  
 respectively, $\lambda_R$.                                                     
 One combination is conventionally fixed in the Higgs phase by the              
 ratio of masses $R_{HW} \equiv M_H/M_W$ as                                     
\be   \label{eq6.1}                                                             
R_{HW}^2 = \frac{32\lambda_R}{g_R^2} \ .                                        
\ee                                                                             
 This corresponds to the relations of the masses to the renormalized            
 vacuum expectation value: $M_H^2=8\lambda_R v_R^2$, respectively,              
 $M_W^2=g_R^2 v_R^2/4$.                                                         
 For the correct normalization of $\lambda_R$ one has to remember               
 that it corresponds to the bare coupling in perturbation theory                
 $\lambda_0$, which is related to $\lambda$ by                                  
\be   \label{eq6.2}                                                             
\lambda_0 \equiv \frac{\lambda}{4\kappa^2}                                      
\ee                                                                             
 and $\lambda_0$ is often multiplied by $24=4!$.                                
                                                                                
 Note that for fixed gauge coupling $g_R$, because of the                       
 Weinberg-Linde bound on the Higgs boson mass \cite{WEILIN},                    
 $R_{HW}$ and $\lambda_R$ defined by (\ref{eq6.1}) have positive lower          
 bounds.                                                                        
                                                                                
 The mass ratio $R_{HW}$ has been determined in the previous section            
 (see table \ref{tab5.2}).                                                      
 Now we shall introduce the renormalized gauge coupling  $g_R$ in a             
 way which is convenient for numerical simulations.                             
                                                                                
%%%%%%%%%%%%%%%%%%%%%%%%%%%%%%%%%%%%%%%%%%%%%%%%%%%%%%%%%%%%%%%%%%%%%%%%        
\subsection{Renormalized gauge coupling}             \label{sec6.1}             
\begin{figure}                                                                  
\vspace{9.0cm}                                                                  
\includegraphics{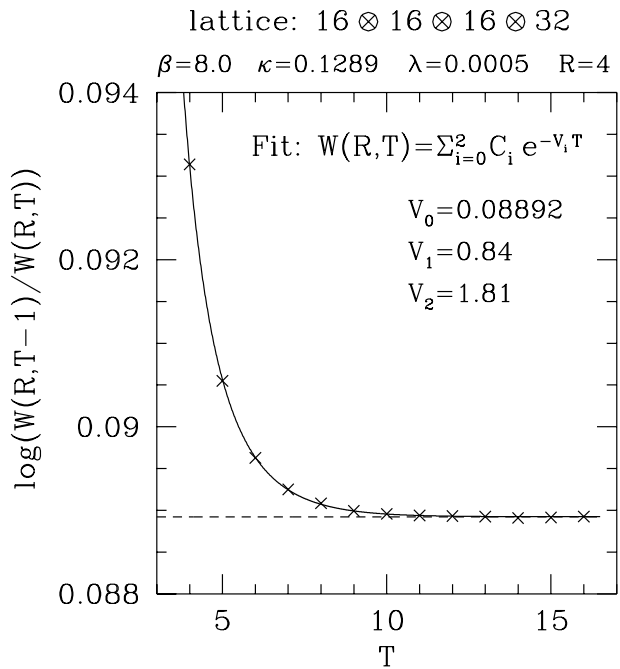}                                                
\begin{center}                                                                  
\parbox{15cm}{\caption{ \label{fig6.1}                                          
 The $T$-dependence of the Wilson loop ratio $W(R,T-1)/W(R,T)$                  
 in the $h2[16/90]$ point at $R=4$.                                             
 The curve is the best fit with three exponentials.                             
}}                                                                              
\end{center}                                                                    
\end{figure}

 The static potential at distance $R$ can be obtained from the                  
 Wilson loops $W(R,T)$ by                                                       
\be                                                                             
V(R)\equiv-\lim_{T \rightarrow \infty}\frac{1}{T}\log W(R,T) \ .                
\ee                                                                             
 We determined in the points defined by table \ref{tab5.1} the on-axis          
 Wilson loops of size $R\otimes T$ with $1\leq R\leq L_s/2$ and                 
 $1\leq T\leq L_t/2$ on $L_s^3 \cdot L_t$ lattices.                                      
 Every rectangular Wilson loop with two sides in time direction                 
 was included in the statistics.                                                
 These Wilson loops were evaluated after transforming the gauge                 
 configuration to temporal gauge.                                               
 The time dependence has been fitted by three exponentials in order             
 to determine the large time asymptotics.
 We checked that by leaving out one exponential and going to
 larger time distances the lowest energy $V_0$ does not change.                                       
 In figure \ref{fig6.1} we show a typical example for the behaviour of          
 the logarithm of the ratio  $W(R,T-1)/W(R,T)$.                                 
%%%%%%%%%%%%%%%%%%%%%%%%%%%%%%%%%%%%%%%%%%%%%%%%%%%%%%%%%%%%%%%%%%%%%%%%        
%%%%%%%%%%%%%%%%%%%%%%%%%%%%%%%%%%%%%%%%%%%%%%%%%%%%%%%%%%%%%%%%%%%%%%%%        
 As it can been seen from the given fit parameters,                             
 the ground state and the excited states are separated by a large               
 energy gap.                                                                    
 The situation for all other points is quite similar.                           
 Due to this, there is no need for an optimization of the source                
 to enhance the ground state signal.                                            
                                                                                
\begin{figure}                                                                  
\vspace{9.0cm}                                                                  
\includegraphics{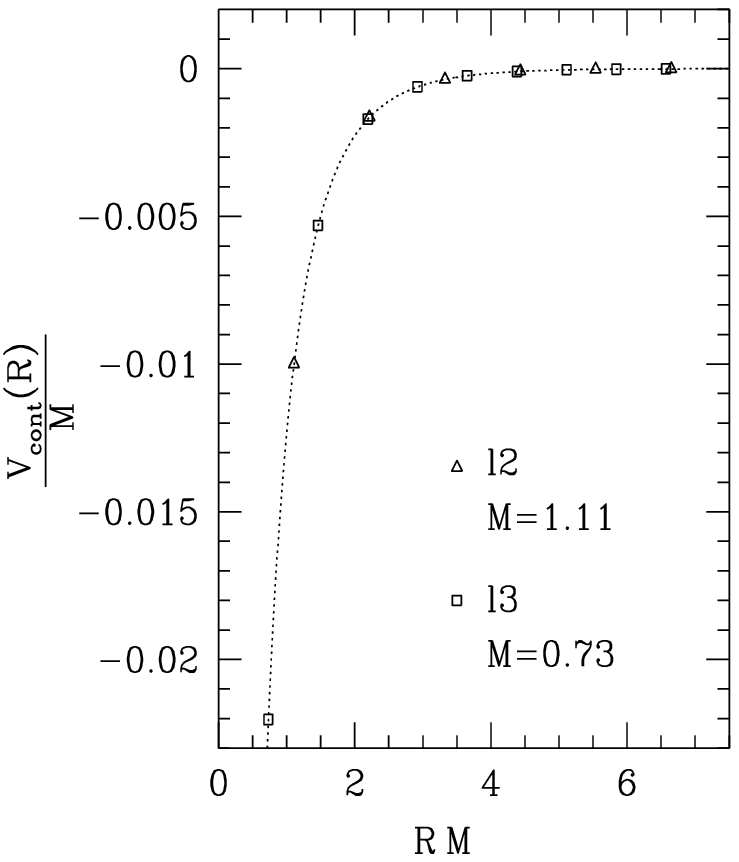}                                                
\includegraphics{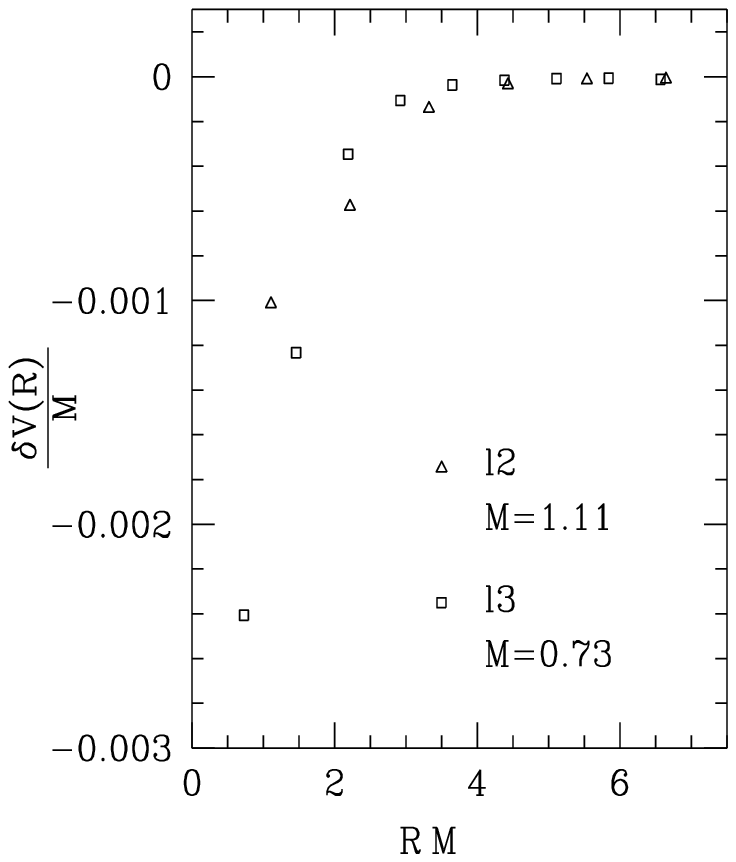}                                               
\begin{center}                                                                  
\parbox{15cm}{\caption{ \label{fig6.2}                                          
 The continuum potential extracted (left picture) and the lattice               
 correction term (right picture) in points $l2$ and $l3$ as a function          
 of distance.                                                                   
 $M$ is the screening mass.                                                     
 The dotted curve is the best fit for $l3$.                                     
}}                                                                              
\end{center}                                                                    
\end{figure}

 The static potential can be fitted well by a Yukawa-term with lattice          
 corrections, as discussed in \cite{LAMOWE}.                                    
 It takes the form                                                              
\be \label{eq6.3}                                                               
V(R)=-\frac{A}{R} e^{-M R}+C+D\, G(M,R,L_s) \ .                                 
\ee                                                                             
 The parameter $M$ is the {\em screening mass} which is closely related         
 but not exactly equal to the physical W-boson mass $M_W$ determined            
 from the correlation functions in section \ref{sec5.1}.                        
 $G(M,R,L_s)$ describes the difference between the continuum potential          
 and the finite lattice version to lowest order.                                
 For on-axis $R$ values it is given by                                          
\be                                                                             
G(M,R,L_s)=\frac{1}{R}e^{-M R}-I(M,R,L_s)  \ ,                                      
\ee                                                                             
where                                                                           
\be                                                                             
I(M,R,L_s) \equiv\frac{4\pi}{L_s^3}                                             
\sum_{k_i} \frac{e^{ik_3R}}{M^2+4\sum_{i=1}^{3}\sin^2(k_i/2)} \ ,               
\ee                                                                             
\be                                                                             
k_i=\frac{2\pi n_i}{L_s} \ , \hspace{3em} n_i=0,\ldots,L_s-1 \ .                
\ee                                                                             
 In figure \ref{fig6.2} we show the continuum potential                         
 $V_{cont}(R)=V(R)-C-D\,G(M,R,L_s)$ and the correction term                     
 $\delta V(R)=D\,G(M,R,L_s)$ for the two points $l2$ and $l3$.                  
%%%%%%%%%%%%%%%%%%%%%%%%%%%%%%%%%%%%%%%%%%%%%%%%%%%%%%%%%%%%%%%%%%%%%%%%        
%%%%%%%%%%%%%%%%%%%%%%%%%%%%%%%%%%%%%%%%%%%%%%%%%%%%%%%%%%%%%%%%%%%%%%%%        
 One can see that on the $L_t=3$ lattice (point $l3$) the lattice               
 artifacts are at larger physical distances smaller but at $R=1,2$              
 larger than on the $L_t=2$ lattice (point $l2$).                               
                                                                                
 The renormalized gauge coupling can be defined from the parameters             
 of the static potential as $g_R^2=16\pi A/3$.                                  
 In table \ref{tbpotential} we collected the best fit parameters                
 together with this global definition of $g_R^2$.                               
 Another local definition of $g_R^2$, which is shown in the last                
 column, will be described below.                                               
%%%%%%%%%%%%%%%%%%%%%%%%%%%%%%%%%%%%%%%%%%%%%%%%%%%%%%%%%%%%%%%%%%%%%%%         
\begin{table}                                                                   
\begin{center}                                                                  
\parbox{15cm}{\caption{ \label{tbpotential}                                     
 Summary of the fit parameters for the static potential and the                 
 renormalized gauge coupling for both definitions.                              
}}                                                                              
\end{center}                                                                    
\begin{center}                                                                  
\begin{tabular}{|c||l|l|l|l|l||l|}                                              
\hline                                                                          
 index   & \multicolumn{1}{c|}{$A$} & \multicolumn{1}{c|}{$M$}  &               
           \multicolumn{1}{c|}{$D$} & \multicolumn{1}{c|}{$C$}  &               
           \multicolumn{1}{c||}{$g_R^2 \equiv {16 \over 3}\pi A$} &             
           \multicolumn{1}{c|}{$g_R^2(M^{-1})$}                 \\              
\hline\hline                                                                    
$ l2 $  & 0.0337(3)  &  1.113(16)  &                                            
  0.0344(23)   &  0.066453(6)  &  0.564(6) & 0.563(6) \\                        
\hline                                                                          
$ l3 $  & 0.03340(6)  & 0.730(4) &                                              
  0.0347(12)  & 0.078005(3) & 0.5597(10) &  0.5611(23)     \\                   
\hline                                                                          
$ h2[12] $  &  0.03434(7)  &  0.4273(21)  &                                     
  0.0352(8)   &  0.090768(18)  &  0.5754(11) & 0.5782(25)  \\                   
\hline                                                                          
$ h2[16/85] $  &  0.03440(6) &  0.4130(12)  &                                   
  0.0365(8)  &  0.091401(5)  & 0.5763(10)  & 0.5822(17)    \\                   
\hline                                                                          
$ h2[16/90] $  &  0.03431(3)  &  0.4357(7)  &                                   
  0.0373(4)  &  0.090571(3)  &  0.5749(5)   &  0.5829(11)  \\                   
\hline                                                                          
$ h3 $  &  0.03373(8)  &  0.2538(15)  &                                         
  0.0366(9)  &  0.095584(19)  &  0.5651(13)  &  0.570(7)   \\                   
\hline                                                                          
\end{tabular}                                                                   
\end{center}                                                                    
\end{table}                                                                     
%%%%%%%%%%%%%%%%%%%%%%%%%%%%%%%%%%%%%%%%%%%%%%%%%%%%%%%%%%%%%%%%%%%%%%%         
 Obviously the renormalization effect on bare coupling $g^2=0.5$ is             
 moderate.                                                                      
 The nearby equality of $A$ and $D$ shows that the                              
 potential is dominated by the one vector boson exchange \cite{MICHAEL}.        
 One can see also that finite volume effects are small                          
 (points $h2[12]$ and $h2[16/90]$) and that a small change in                   
 $\kappa$ (points $h2[16/85]$ and $h2[16/90]$) leaves $g_R^2$ almost            
 constant.                                                                      
                                                                                
 Another way to define the renormalized gauge coupling was discussed            
 in \cite{LAMOWE}.                                                              
 We use a slightly different version which is similar to the                    
 procedure recently proposed for pure gauge theory \cite{SOMMER}.               
 Given the potential, the renormalized coupling at distance $R_I$ is            
 defined by                                                                     
\be \label{eq6.4}                                                               
g_R^2(R_I)\equiv \frac{16\pi}{3}                                                
\frac{V(R)-V(R-1)}{I(M,R-1,L_s)-I(M,R,L_s)}  \ .                                
\ee                                                                             
 This can be interpolated to a physical distance, for instance                  
 $r=M^{-1}$.                                                                    
 $R_I$ is defined by the force as the solution of the equation                  
\be                                                                             
\frac{1}{R_I} e^{-M R_I} \left [ \frac{1}{R_I}+M \right ]                       
= I(M,R-1,L_s)-I(M,R,L_s) \ .                                                   
\ee                                                                             
 The value of $M$ in eq.~(\ref{eq6.4}) can be taken from the fit               
 (\ref{eq6.3}) restricted to large distances.                                   
 The reason of introducing a local definition of $g_R$ is that the short        
 distance potential is not purely Yukawa like.                                  
 Specifying the distance is therefore in principle important.                   
 However, our resolution is up to now not fine enough to see the                
 logarithmic dependence on distance.                                            
                                                                                
 The time dependence of the Wilson loops as well as the distance                
 dependence of the static potential were described very well by the             
 ans\"{a}tze, i.e.\ $\chi^2$ was of order one in both cases.                    
 The statistical errors were determined from independent subsamples             
 (see previous section).                                                        
 The error for the renormalized gauge coupling squared $g_R^2(M^{-1})$          
 quoted in table \ref{tbpotential} is the sum of the statistical errors         
 for potential and mass, and the systematical error.                            
 The interpolation to the distance $M^{-1}$ was done by a linear                
 approximation of the two neighbouring points.                                  
 The systematical errors are estimated from the change arising by the           
 inclusion of a third point.                                                    
 The given errors are usually dominated by the statistical errors of $M$.               
 Note that the value of $g_R^2(M^{-1})$ in point $l2$ is corrected              
 as compared to \cite{CFHJJM}.                                                  
                                                                                
%%%%%%%%%%%%%%%%%%%%%%%%%%%%%%%%%%%%%%%%%%%%%%%%%%%%%%%%%%%%%%%%%%%%%%%%        
\subsection{Renormalization group trajectories}      \label{sec6.2}             
 The continuum limit of quantum field theories on the lattice is taken          
 along renormalization group trajectories, also called {\em lines of            
 constant physics} (LCP's).                                                     
 Going to the continuum limit along such lines in bare parameter space          
 the same physical theory is reproduced with increasing precision.              
 This means that the physical results of numerical simulations should           
 be the same, apart from ``scale breaking lattice artifacts''.                  
 In the SU(2) Higgs model, in order to define the LCP's, one has to             
 keep fixed the values of two independent renormalized couplings,               
 say the above discussed renormalized quartic ($\lambda_R$) and                 
 gauge ($g_R$) couplings.                                                       
 The third parameter characterizing the points along LCP's can be               
 chosen as                                                                      
\be   \label{eq6.5}                                                             
\tau \equiv \log M_W^{-1}  \  .                                                 
\ee                                                                             
 (Remember that in this paper the lattice spacing is set to $a=1$,              
 therefore $M_W$ is measured here in lattice units.)                            
%%%%%%%%%%%%%%%%%%%%%%%%%%%%%%%%%%%%%%%%%%%%%%%%%%%%%%%%%%%%%%%%%%%%%%%%        
\begin{figure}                                                                  
\vspace{9.0cm}                                                                  
\includegraphics{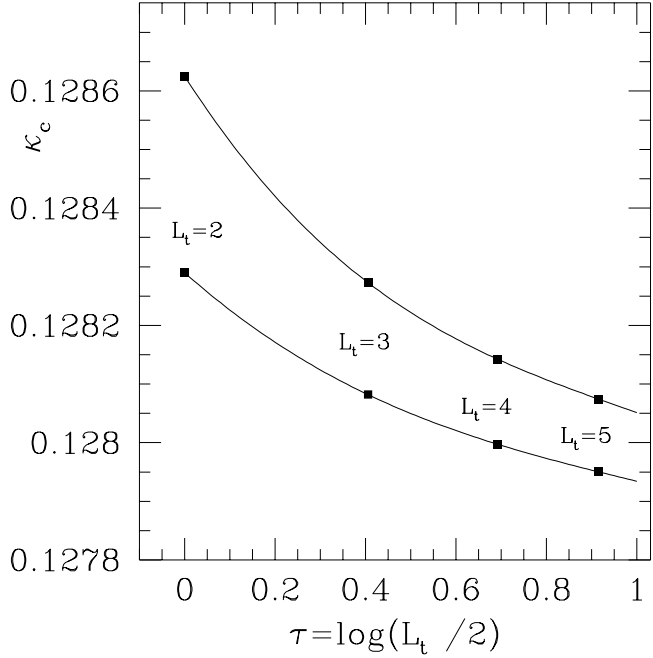}                                                 
\begin{center}                                                                  
\parbox{15cm}{\caption{ \label{fig6.3}                                          
 The positions of the phase transition in $\kappa$ along two lines of           
 constant physics, which start at the phase transition points of an             
 $L_t=2$ lattice at $\beta=8,\;\lambda=0.0001$ (lower curve) and                
 $\beta=8,\;\lambda=0.0005$ (upper curve).                                      
 The points with $2 \leq L_t \leq 5$ are determined from the one-loop           
 invariant effective potential.                                                 
 The curves are polynomial interpolations.                                      
}}                                                                              
\end{center}                                                                    
\end{figure}                                                                    
%%%%%%%%%%%%%%%%%%%%%%%%%%%%%%%%%%%%%%%%%%%%%%%%%%%%%%%%%%%%%%%%%%%%%%%%        
                                                                                
 Since our numerical simulations are performed in the weak coupling             
 region of parameter space, the change of the bare couplings                    
 $g^2 \equiv 4/\beta$ and $\lambda_0$ defined by eq.~(\ref{eq6.2})             
 along LCP's can be well approximated by the solutions of the                   
 one-loop perturbative renormalization group equations:                         
$$                                                                              
\frac{dg^2(\tau)}{d\tau} = \frac{1}{16\pi^2} \left[                             
- \frac{43}{3}g^4                                                               
+ {\cal O}(\lambda_0^3,\lambda_0^2g^2,\lambda_0g^4,g^6)                         
\right] \ ,                                                                     
$$                                                                              
\be   \label{eq6.6}                                                             
\frac{d\lambda_0(\tau)}{d\tau} = \frac{1}{16\pi^2} \left[                       
96\lambda_0^2 + \frac{9}{32}g^4 - 9\lambda_0 g^2 +                              
{\cal O}(\lambda_0^3,\lambda_0^2g^2,\lambda_0g^4,g^6) \right] \ .               
\ee                                                                             
 The integration of these equations can be started at the transition            
 points of the $L_t=2$ lattices.                                                
 At a distance $\Delta\tau=\log(L_t/2)$ ($L_t=3,4,5,\ldots$) in                 
 parameter $\tau$ one obtains the corresponding values of                       
 $(g^2,\lambda_0)$ for the transition points of lattices with temporal          
 extension $L_t$.                                                               
                                                                                
 In order to have a perturbative prediction also for the change of              
 the third bare parameter $\kappa$, one can make use of the one-loop            
 perturbative invariant effective potential.                                    
 The infinite volume predictions for our ``low'' and ``high'' values of         
 the quartic coupling in the range $2 \leq L_t \leq 5$ are shown by             
 fig.~\ref{fig6.3}.                                                            
 As we discussed in section \ref{sec4}, the measured values of                  
 $\kappa_c$ are very well reproduced by the one-loop invariant                  
 effective potential for $L_t=2,3$ at our ``low'' value of the quartic          
 coupling.                                                                      
 At the ``high'' quartic coupling there is a parallel shift between             
 prediction and measurements: the prediction is too low by                      
 $\Delta\kappa_c = 0.00025(1)$.                                                 
 As a consequence, the derivatives of $\kappa$ along these LCP's                
 can be well determined:                                                        
$$                                                                              
\frac{\partial\kappa^{l2}}{\partial\tau} = -0.000685(40+69) \ ,                 
\hspace{3em}                                                                    
\frac{\partial\kappa^{h2}}{\partial\tau} = -0.00120(9+5)  \ ,                   
$$                                                                              
\be   \label{eq6.7}                                                             
\frac{\partial\kappa^{l3}}{\partial\tau} = -0.000367(17+4) \ ,                  
\hspace{3em}                                                                    
\frac{\partial\kappa^{h3}}{\partial\tau} = -0.000588(35+6) \ .                  
\ee                                                                             
 The errors given here include an estimate of the systematic                    
 error by comparing different polynomial interpolations of the points           
 $2 \leq L_t \leq 5$ (first entry in parentheses) plus the error                
 coming from the uncertainties of $\kappa_c$ obtained in the                    
 numerical simulations (second entry).                                          
                                                                                
 Although the estimates of the LCP's by one-loop perturbation theory            
 are very useful for guiding the numerical simulations, finally                 
 one has to confront them with numerical simulation data.                       
 Having this in mind we have chosen the $L_t=3$ points with index               
 $l3$ and $h3$ on the trajectories of the solutions                             
 of eq.~(\ref{eq6.6}) which start at $l2$ and $h2$, respectively.              
 (For $\beta$ and $\lambda$ we kept only the first few digits which             
 are sufficient within our typical statistical errors.)                         
 Comparing the values of $R_{HW}$ and $g_R^2$ in the points $l2$ and            
 $l3$, respectively, $h2$ and $h3$ (see tables \ref{tab5.2} and                 
 \ref{tbpotential}) one can conclude that the one-loop predictions from         
 eq.~(\ref{eq6.6}) are very good: the measured values of both                  
 renormalized couplings in corresponding points are very close to               
 each other, in most cases equal within errors.                                 
                                                                                
%%%%%%%%%%%%%%%%%%%%%%%%%%%%%%%%%%%%%%%%%%%%%%%%%%%%%%%%%%%%%%%%%%%%%%%%        
                                                                                
\section{Latent heat}                          \label{sec7}                     
 An important characteristic feature of first order phase transitions           
 is the latent heat, i.e.\ the discontinuity of the energy density              
 $\epsilon$.                                                                    
 The pressure $P$ is continuous, therefore $\Delta\epsilon$ can                 
 be obtained from the discontinuity of $\delta \equiv \epsilon/3-P$.            
 This latter quantity is related to the trace of the energy-momentum            
 tensor and is theoretically simple, because it can be obtained from            
 the derivative of the action density with respect to the lattice               
 spacing $a$.                                                                   
 This, in turn, implies that it is closely related to the LCP's                 
 for the continuum limit discussed in the previous section.                     
                                                                                
 As it has been shown in \cite{CFHJJM}, the latent heat                         
 $\Delta\epsilon$ in the SU(2) Higgs model is given by                          
\be \label{eq7.1}                                                               
\frac{\Delta\epsilon}{T_c^4} = L_t^4                                            
\left\langle \frac{\partial\kappa}{\partial\tau}                                
\cdot 8\Delta L_{\varphi,x\mu}                                                  
- \frac{\partial\lambda}{\partial\tau} \cdot \Delta Q_x                         
- \frac{\partial\beta}{\partial\tau}                                            
\cdot 6\Delta P_{pl} \right\rangle \ .                                          
\ee                                                                             
 Here, besides the notations in (\ref{eq5.2}), we used                          
\be \label{eq7.2}                                                               
Q_x \equiv (\rho_x^2 - 1)^2 \ ,                                                 
\hspace{3em}                                                                    
P_{pl} \equiv 1 - \half {\rm Tr\,} U_{pl} \ .                                   
\ee                                                                             
 The sign of the discontinuities like $\Delta\epsilon$ etc.\ will               
 be defined in such a way that they are differences of values in                
 the symmetric phase minus Higgs phase.                                         
                                                                                
 The obtained average values of the global quantities in eqs.                   
 (\ref{eq5.1}), (\ref{eq5.2}), (\ref{eq7.2}) and of the                         
 lattice action per point                                                       
\be \label{eq7.3}                                                               
S_x \equiv 6\beta P_{pl} + R_x + \lambda Q_x - 8\kappa L_{\varphi,x\mu}         
\ee                                                                             
 are shown in table \ref{tab7.1}.                                               
%%%%%%%%%%%%%%%%%%%%%%%%%%%%%%%%%%%%%%%%%%%%%%%%%%%%%%%%%%%%%%%%%%%%%%%%        
\begin{table}                                                                   
\begin{center}                                                                  
\parbox{15cm}{\caption{ \label{tab7.1}                                          
 The obtained values of global averages ``above'' and ``below''                 
 phase transition points, i.e.\ in the Higgs, respectively,                     
 symmetric phase.                                                               
}}                                                                              
\end{center}                                                                    
\begin{center}                                                                  
\begin{tabular}{|c||c|c|c|c|c|c|}                                               
\hline                                                                          
index & $P_{pl}$  &  $R_x$  &  $L_{\alpha,x\mu}$  &                             
$L_{\varphi,x\mu}$  &  $Q_x$  &  $S_x$         \\                               
\hline\hline                                                                    
$l2[a]$  &                                                                      
0.085487(5) & 25.184(12)  & 0.91666(4)  & 22.720(12)  &  652(2)     &           
6.0333(3)   \\                                                                  
\hline                                                                          
$l2[b]$  &                                                                      
0.096053(1) & 2.88354(17) & 0.27614(3)  & 0.86267(16) &  7.5(1)     &           
6.60939(2)  \\                                                                  
\hline                                                                          
$l3[a]$  &                                                                      
0.09009(12) & 11.548(24)  & 0.8284(3)   & 9.349(24)   & 138.4(5)    &           
6.3878(6)   \\                                                                  
\hline                                                                          
$l3[b]$  &                                                                      
0.094438(1) & 2.6585(3)   & 0.21724(7)  & 0.6442(3)   & 6.2802(17)  &           
6.61696(3)  \\                                                                  
\hline                                                                          
$h2[a]$  &                                                                      
0.095391(6) & 3.988(11)   & 0.4629(15)  & 1.947(11)   & 16.17(9)    &           
6.5677(3)   \\                                                                  
\hline                                                                          
$h2[b]$  &                                                                      
0.096000(1) & 2.9761(15)  & 0.2964(3)   & 0.9568(15)  & 8.302(10)   &           
6.60189(6)  \\                                                                  
\hline                                                                          
$h3[a]$  &                                                                      
0.094183(2) & 3.133(3)    & 0.3243(6)   & 1.113(3)    & 9.31(2)     &           
6.59861(9)  \\                                                                  
\hline                                                                          
$h3[b]$  &                                                                      
0.094421(1) & 2.7006(11)  & 0.2290(3)   & 0.6896(11)  & 6.517(7)    &           
6.61214(4)  \\                                                                  
\hline                                                                          
\end{tabular}                                                                   
\end{center}                                                                    
\end{table}                                                                     
%%%%%%%%%%%%%%%%%%%%%%%%%%%%%%%%%%%%%%%%%%%%%%%%%%%%%%%%%%%%%%%%%%%%%%%%        
 The indices of points where the numerical simulations were                     
 performed are defined in table \ref{tab5.3}.                                   
                                                                                
 As it has been discussed above (see e.g.~section \ref{sec5.2}),               
 the small errors in table \ref{tab7.1} could be achieved by avoiding           
 errors of extrapolation.                                                       
 This is made possible by performing the simulations on large enough            
 lattices, where the strong metastability of phases can be exploited.           
 In this way the contribution of the statistical errors in table                
 \ref{tab7.1} to the errors of $\Delta\epsilon$ is negligibly                   
 small compared to the errors of the derivatives.                               
 In particular, in our case the errors in eq.~(\ref{eq7.4}) are                
 dominated by the errors coming from eq.~(\ref{eq6.7}).                        
                                                                                
 Besides the discontinuities of global quantities, for                          
 $\Delta\epsilon$ in eq.~(\ref{eq7.1}) the derivatives of bare                 
 parameters along the LCP's are needed.                                         
 These have been discussed in the previous section.                             
 $\partial\beta/\partial\tau$ and $\partial\lambda/\partial\tau$                
 can be obtained to a good approximation from eq.~(\ref{eq6.6}).               
 The values of $\partial\kappa/\partial\tau$ have also been evaluated           
 from the one-loop invariant effective potential and from numerical             
 simulation data.                                                               
 They are given by eq.~(\ref{eq6.7}).                                          
 Inserting them into eq.~(\ref{eq7.1}) together with the numbers               
 from table \ref{tab7.1} one obtains                                            
$$                                                                              
\left( \frac{\Delta\epsilon}{T_c^4} \right)^{l2} = 1.81(29)   \ ,               
\hspace{3em}                                                                    
\left( \frac{\Delta\epsilon}{T_c^4} \right)^{h2} = 0.132(17)  \ ,               
$$                                                                              
\be \label{eq7.4}                                                               
\left( \frac{\Delta\epsilon}{T_c^4} \right)^{l3} = 1.57(13)   \ ,               
\hspace{3em}                                                                    
\left( \frac{\Delta\epsilon}{T_c^4} \right)^{h3} = 0.122(9)   \ .               
\ee                                                                             
 We see from here that within errors the results on $L_t=2$                     
 and $L_t=3$ lattices coincide.                                                 
                                                                                
 Comparing the contributions of different terms to $\Delta\epsilon$             
 in (\ref{eq7.1}) one sees that the term with $L_\varphi$                       
 dominates.                                                                     
 This is in accordance with the observation that the lattice                    
 spacing is mainly determined by the hopping parameter $\kappa$                 
 (see sections \ref{sec4} and \ref{sec5}).                                      
                                                                                
%%%%%%%%%%%%%%%%%%%%%%%%%%%%%%%%%%%%%%%%%%%%%%%%%%%%%%%%%%%%%%%%%%%%%%%%        
                                                                                
\section{Interface tension}                    \label{sec8}                     
                                                                                
 At the transition point of the electroweak phase transition                    
 mixed states can appear, where different bulk phases are separated by          
 interfaces.                                                                    
 The interface tension, $\sigma$, is the free energy density associated         
 with these interfaces.                                                         
 The dynamics of a first order phase transition is to a large extent            
 determined by the interface tension.                                           
 In particular, the nucleation of bubbles or droplets is essentially            
 influenced by it \cite{LANGER,CSEKAP}.                                         
 In perturbation theory the determination of $\sigma$ is a task                 
 which is far from being trivial.                                               
 The main problem is the bad infrared behaviour of the finite                   
 temperature effective action in the $SU(2)$-Higgs model.                       
 Therefore, non-perturbative lattice simulations are very useful.               
                                                                                
 In this section we use two different methods, namely, the two-coupling         
 method and Binder's histogram method combined with multicanonical              
 updating, to determine $\sigma$.                                               
 In both cases we use elongated lattices.                                       
 This geometric choice of the lattices results in interfaces which are          
 perpendicular to the long direction.                                           
 Other configurations (e.g.~multiple walls, large spherical bubbles)           
 are highly suppressed.                                                         
 The inverse temperature will be restricted in this section to $L_t=2$.         
                                                                                
%%%%%%%%%%%%%%%%%%%%%%%%%%%%%%%%%%%%%%%%%%%%%%%%%%%%%%%%%%%%%%%%%%%%%%%%        
\subsection{Two-coupling method and interface tension}\label{sec8.1}            
 One way to determine the interface tension in the SU(2) Higgs model            
 is based on a modified version of the Potvin-Rebbi two-coupling method         
 \cite{POTREB}.                                                                 
 Similarly to subsection \ref{sec4.2}, a lattice is considered with             
 a long extension in the $z$-direction.                                         
 The long direction with periodic boundary conditions is divided into           
 two equal halves with different couplings, in such a way that                  
 an interface pair is created near the hypersurfaces where the                  
 couplings are changing.                                                        
                                                                                
 Since we have three bare parameters, in principle one can choose               
 any of them (or some combination) to be different in the two halves            
 of the lattice.                                                                
 A physically good choice would be to change the temperature only.              
 One can move, for instance, along an LCP (see section \ref{sec6.2})            
 when only the lattice spacing $a$ is changing.                                 
 Assuming that the volume is very large, the only relevant change is            
 in the temperature $T=1/(aL_t)$.                                               
 A simpler way, which comes close to this, is to change only $\kappa$.          
 The reason is that in the latent heat the contribution of the                  
 $\varphi$-link $L_\varphi$ dominates which is conjugate to the                 
 hopping parameter $\kappa$.                                                    
                                                                                
 Therefore we decided to split the value of $\kappa$.                           
 On an $L_t \cdot L_x \cdot L_y \cdot L_z$ lattice with                         
 $L_z \gg L_{x,y,t}$, and in our case $L_t=2$, $L_x=L_y \equiv L_{xy}$,         
 the two halves in the $z$-direction have hopping parameters                    
 $\kappa_{1,2}$ which are slightly below and above the transition point:        
 $\kappa_1 < \kappa_c < \kappa_2$.                                              
                                                                                
 The interface tension is defined by the ``free energy'' (more                  
 precisely ``free action'')                                                     
\be \label{eq8.1}                                                               
F \equiv S - \Omega d(s) \equiv \Omega f(s) \ ,                                 
\ee                                                                             
 where $S$ is the lattice action (\ref{eq1.1}),                                 
 $\Omega \equiv L_x L_y L_z L_t$ is the number of lattice points and            
 $d(s)$ is the spectral density of states as a function of the                  
 action density $s \equiv S/\Omega$:                                            
\be \label{eq8.2}                                                               
e^{\Omega d(s)} \equiv                                                          
\int [dU] [d\varphi] \delta\left( s - \frac{S}{\Omega} \right) \ .              
\ee                                                                             
 This implies that the probability distribution of $s$ is                       
 $w(s) \propto \exp[-\Omega f(s)]$.                                             
 The interface tension between the states with $\kappa_1$ and                   
 $\kappa_2$ is given in lattice units by                                        
$$                                                                              
\sigma = (2L_x L_y L_t)^{-1}                                                    
\left[ F(\kappa_1,\kappa_2) - \half F(\kappa_1,\kappa_1)                        
- \half F(\kappa_2,\kappa_2) \right]                                            
$$                                                                              
\be \label{eq8.3}                                                               
= \frac{1}{4L_x L_y L_t} \left\{                                                
\left[ F(\kappa_1,\kappa_2) -  F(\kappa_1,\kappa_1) \right] -                   
\left[ F(\kappa_2,\kappa_2) -  F(\kappa_1,\kappa_2) \right]                     
\right\} \ .                                                                    
\ee                                                                             
 Let us write the action on the lattice with two halves as                      
\be \label{eq8.4}                                                               
S \equiv S_0 - \kappa_1 S_1 - \kappa_2 S_2 \ ,                                  
\ee                                                                             
 where $S_0$ is the piece independent from $\kappa$.                            
 Then in the thermodynamical limit we have for $i=1,2$                          
\be \label{eq8.5}                                                               
\frac{\partial F(\kappa_1,\kappa_2)}{\partial\kappa_i}                          
= -\langle S_i \rangle_{\kappa_1,\kappa_2} \ ,                                  
\ee                                                                             
 if $\langle \ldots \rangle_{\kappa_1,\kappa_2}$ denotes                        
 expectation values for a lattice with two halves at $\kappa_{1,2}$.            
 Therefore we obtain from (\ref{eq8.3})                                         
\be \label{eq8.6}                                                               
\sigma(\kappa_1,\kappa_2) = (4L_x L_y L_t)^{-1} \left\{                         
\int_{\kappa_1}^{\kappa_2} d\kappa                                              
\langle S_1 \rangle_{\kappa,\kappa_2} -                                         
\int_{\kappa_1}^{\kappa_2} d\kappa                                              
\langle S_2 \rangle_{\kappa_1,\kappa}  \right\} \ .                             
\ee                                                                             
 Here the notation $\sigma(\kappa_1,\kappa_2)$ emphasizes the dependence        
 on $\kappa_{1,2}$.                                                             
 Finally, since $S_{1,2}$ is given by the $\varphi$-link                        
 $L_\varphi$, we have                                                           
\be \label{eq8.7}                                                               
\sigma(\kappa_1,\kappa_2) = L_z \left\{                                         
  \int_{\kappa_1}^{\kappa_2} d\kappa L_\varphi^{(1)}(\kappa,\kappa_2)           
- \int_{\kappa_1}^{\kappa_2} d\kappa L_\varphi^{(2)}(\kappa_1,\kappa)           
\right\} \ ,                                                                    
\ee                                                                             
 where $L_\varphi^{(1,2)}(\kappa,\kappa^\prime)$ denote the expectation         
 values of $L_\varphi$ averaged in the two halves, if the hopping               
 parameters are $\kappa$ and $\kappa^\prime$, respectively.                     
                                                                                
 In order to obtain the physically interesting interface tension one            
 has, of course, to perform the non-commutative limits                          
 $L_{x,y,z} \to \infty$ and                                                     
 $\Delta\kappa \equiv \kappa_2-\kappa_c = \kappa_c-\kappa_1 \to 0$.             
 For a given lattice extension $\Delta\kappa$ cannot be                         
 arbitrarily small, because if the difference in free energy density            
 becomes small tunneling into the other phase can occur and the                 
 interfaces disappear.                                                          
 The presence of the interfaces can, however, be monitored to ensure            
 the applicability of (\ref{eq8.7}).                                            
                                                                                
 For small $\Delta\kappa$ the integrals in (\ref{eq8.7}) can be well            
 approximated by the average of the integrand at the two end points.            
 Besides, for equal arguments we obviously have                                 
 $L_\varphi^{(1)}(\kappa,\kappa) = L_\varphi^{(2)}(\kappa,\kappa)$.             
 This gives                                                                     
\be \label{eq8.8}                                                               
\sigma \simeq L_z \Delta\kappa \left[                                           
 L_\varphi^{(1)}(\kappa_1,\kappa_2) - L_\varphi^{(1)}(\kappa_1,\kappa_1)        
+L_\varphi^{(2)}(\kappa_2,\kappa_2) - L_\varphi^{(2)}(\kappa_1,\kappa_2)        
\right] .                                                                       
\ee                                                                             

 We are interested in the limit $\Delta\kappa \to 0$.                           
 Therefore in the square bracket only contributions of order                    
 $1/\Delta\kappa$ matter.                                                       
 Such contributions cannot come from terms with equal hopping                   
 parameters in the two halves.                                                  
 Therefore let us introduce the parametrizations                                
$$                                                                              
L_\varphi^{(1)}(\kappa_1,\kappa_2) = \frac{-c_1}{\kappa_1-\kappa_c}             
+ b_1 + a_1(\kappa_1-\kappa_c) + {\cal O}(\kappa_1-\kappa_c)^2 \ ,              
$$                                                                              
\be \label{eq8.9}                                                               
L_\varphi^{(2)}(\kappa_1,\kappa_2) = \frac{-c_2}{\kappa_2-\kappa_c}             
+ b_2 + a_2(\kappa_2-\kappa_c) + {\cal O}(\kappa_2-\kappa_c)^2 \ .              
\ee                                                                             
 Then for $\Delta\kappa \to 0$ a finite volume estimator of the                 
 interface tension is                                                           
\be \label{eq8.10}                                                              
\hat{\sigma} = L_z(c_1+c_2) \ .                                                 
\ee                                                                             

 In view of this formula the question arises about the origin                   
 of the contributions of order ${\cal O}(1/\Delta\kappa)$.                      
 For small $\Delta\kappa$ the difference in free energies of the                
 two phases is of the order ${\cal O}(\Delta\kappa)$.                           
 Therefore the interfaces are not fixed at the hypersurfaces                    
 where $\kappa$ changes, but are penetrating into the neighbouring              
 regions with constant $\kappa$.                                                
 If the $\kappa$ change is at $z=z_0$, the probability of the interface         
 position at $z>z_0$ is $\propto\exp[-const.\;\Delta\kappa (z-z_0)]$.           
 Integrating over $z$ gives ${\cal O}(1/\Delta\kappa)$.                         
 In fact, as the numerical simulation data show, the distributions of           
 $z$-slices of $L_\varphi$ as a function of $z$ can be well fitted              
 by exponentials.                                                               
                                                                                
 The above formulas also show that the extrapolation to                         
 $\Delta\kappa=0$ is delicate.                                                  
 The intervals in $\kappa_{1,2}$ for the fits in eq.~(\ref{eq8.9})             
 have to be carefully chosen.                                                   
 These forms are, in fact, replacing (\ref{eq8.7}) in the region where          
 $\sigma(\kappa_1,\kappa_2)$ depends linearly on $\kappa_{1,2}$.                
 For given lattice extensions one cannot take intervals too close               
 to $\kappa_c$.                                                                 
 This is a kind of round-off effects, which usually appear at first             
 order phase transitions in finite volumes but get smaller                      
 for increasing lattice sizes.                                                  
 First, according to eq.~(\ref{eq8.10}) one has to choose                      
 $L_z$ large enough in order that the two interfaces do not                     
 interact with each other.                                                      
 In this case the left hand side becomes independent from $L_z$.                
 Second, for increasing $L_x,\;L_y$ the free energy differences grow            
 proportional to $(L_x L_y\Delta\kappa)$.                                       
 The probability of penetration into the regions with constant                  
 $\kappa$ is exponentially decreasing.                                          
 This allows to consider small $\Delta\kappa$'s of the order of                 
 ${\cal O}(1/(L_x L_y))$.                                                       
                                                                                
 In the numerical simulations with two $\kappa$'s the lattice sizes             
 extended up to $2 \cdot 16^2 \cdot 128$ in the $l2$ point and                  
 $2 \cdot 32^2 \cdot 256$ in the $h2$ point.                                    
 The numerical simulation data on these lattices are collected                  
 in table \ref{tab8.1}, respectively, table \ref{tab8.2}.                       
%%%%%%%%%%%%%%%%%%%%%%%%%%%%%%%%%%%%%%%%%%%%%%%%%%%%%%%%%%%%%%%%%%%%%%%%        
\begin{table}                                                                   
\begin{center}                                                                  
\parbox{15cm}{\caption{ \label{tab8.1}                                          
 The average values of $L_\varphi^{(1,2)}$ obtained in two-$\kappa$             
 simulations with $\kappa=\kappa_{1,2}$ in point $l2$ on                        
 $2 \cdot 16^2 \cdot 128$ lattice.                                              
}}                                                                              
\end{center}                                                                    
\begin{center}                                                                  
\begin{tabular}{|c|c||c|c|}                                                     
\hline                                                                          
$\kappa_1$  &  $\kappa_2$  & $L_\varphi^{(1)}$ & $L_\varphi^{(2)}$  \\          
\hline\hline                                                                    
 0.12790  &  0.12870  &  2.6832(44)  &  53.422(12)    \\  \hline                
 0.12795  &  0.12865  &  2.7148(54)  &  50.088(17)    \\  \hline                
 0.12800  &  0.12860  &  2.7372(65)  &  46.630(16)    \\  \hline                
 0.12805  &  0.12855  &  2.7731(84)  &  43.052(15)    \\  \hline                
 0.12810  &  0.12850  &  2.8323(87)  &{\em 39.300(17)}\\  \hline                
\end{tabular}                                                                   
\end{center}                                                                    
\end{table}                                                                     
%%%%%%%%%%%%%%%%%%%%%%%%%%%%%%%%%%%%%%%%%%%%%%%%%%%%%%%%%%%%%%%%%%%%%%%%        
%%%%%%%%%%%%%%%%%%%%%%%%%%%%%%%%%%%%%%%%%%%%%%%%%%%%%%%%%%%%%%%%%%%%%%%%        
\begin{table}                                                                   
\begin{center}                                                                  
\parbox{15cm}{\caption{ \label{tab8.2}                                          
 The average values of $L_\varphi^{(1,2)}$ obtained in two-$\kappa$             
 simulations with $\kappa=\kappa_{1,2}$ in point $h2$ on                        
 $2 \cdot 32^2 \cdot 256$ lattice.                                              
}}                                                                              
\end{center}                                                                    
\begin{center}                                                                  
\begin{tabular}{|c|c||c|c|}                                                     
\hline                                                                          
$\kappa_1$  &  $\kappa_2$  & $L_\varphi^{(1)}$ & $L_\varphi^{(2)}$  \\          
\hline\hline                                                                    
 0.12880  &  0.12894  &  0.9948(20)   &  2.9325(43)   \\  \hline                
 0.12881  &  0.12893  &  0.9991(29)   &  2.7939(65)   \\  \hline                
 0.12882  &  0.12892  &  1.0173(24)   &  2.6688(54)   \\  \hline                
 0.12883  &  0.12891  &  1.0269(54)   &  2.5217(86)   \\  \hline                
 0.12884  &  0.12890  &  1.0531(85)   &  2.377(10)    \\  \hline                
 0.12885  &  0.12889  &  1.0873(92)   &  2.237(13)    \\  \hline                
 0.12886  &  0.12888  &{\em 1.043(11)}&  1.998(28)    \\  \hline                
\end{tabular}                                                                   
\end{center}                                                                    
\end{table}                                                                     
%%%%%%%%%%%%%%%%%%%%%%%%%%%%%%%%%%%%%%%%%%%%%%%%%%%%%%%%%%%%%%%%%%%%%%%%        
 Fits with the parameters $a,b,c$ in eq.~(\ref{eq8.9}) were                    
 performed, omitting quadratic and higher order terms in $\Delta\kappa$.        
 For the transition points $\kappa_c=0.12830$ and $\kappa_c=0.12887$            
 were assumed in points $l2$ and $h2$, respectively.                            
 The data entries in the tables with {\em italic} were not taken                
 into account in the fits, in order to obtain acceptable $\chi^2$'s.            
 The other points can be well fitted and the $\chi^2$'s turn out                
 to be of the order of the number of degrees of freedom.                        
 Inserting the results into eq.~(\ref{eq8.10}) and assuming                    
 $T_c=1/2$ we obtain                                                            
\be \label{eq8.11}                                                              
\left( \frac{\hat{\sigma}}{T_c^3} \right)^{l2} = 0.84(16) \ ,                   
\hspace{3em}                                                                    
\left( \frac{\hat{\sigma}}{T_c^3} \right)^{h2} = 0.008(2) \ .                   
\ee                                                                             
 The errors here also contain some subjective estimates of the                  
 systematic errors coming from the choice of fit intervals.                     
 The statistical part of the errors was estimated by repeating                  
 the fits with normally distributed input data.                                 
                                                                                
 Concerning the volume dependence of $\hat{\sigma}$ we also collected           
 data in point $l2$ on $2 \cdot 8^2 \cdot 128$ lattice.                         
 Within statistical errors of about 10\% no deviation from the result           
 given in (\ref{eq8.11}) could be detected.                                     
                                                                                
%%%%%%%%%%%%%%%%%%%%%%%%%%%%%%%%%%%%%%%%%%%%%%%%%%%%%%%%%%%%%%%%%%%%%%%%        
\subsection{Multicanonical method and interface tension}\label{sec8.2}          
 In this subsection we present our results for the interface                    
 tension obtained by the histogram method \cite{BINDER}                         
 at the {\it low} point, $L_t=2$.                                               
                                                                                
 At $\kappa_c$ the probability distribution of an order parameter               
 (e.g.~action density $s_{log} \equiv S_{log}/\Omega$ defined by               
 (\ref{slog}), or link variable $L_\varphi$ in (\ref{eq5.2})) develops          
 two peaks.                                                                     
 They correspond to pure phases and the suppressed configurations               
 between the peaks are dominated by mixed states where the phases               
 are separated by interfaces.                                                   
 Defining $\kappa_c$ by the equal height signal the suppression                 
 at infinite volume is given by the interface tension                           
\be                                                                             
\sigma_\infty=                                                                  
\lim_{\Omega \rightarrow \infty} \sigma_\Omega \ ,                              
\hspace{4em}                                                                    
\sigma_\Omega={1 \over 2 L_x L_y L_t} \log{p_{max} \over p_{min}} \ ,           
\ee                                                                             
 where $p_{max}$ corresponds to the heights of the peaks and                    
 $p_{min}$ to the minimum in between.                                           
                                                                                
 Recently intensive studies have been carried out in order to                   
 understand the finite size corrections to $\sigma$                             
 \cite{BUNK,WIESE,CASELLE1,IWASAKI,CASELLE2}.                                   
 We follow \cite{IWASAKI} and for our simulations at the ``low''                
 value of the quartic coupling $\lambda=0.0001$ use elongated lattices          
 with extensions $\Omega=2\cdot 4^2 \cdot  64$,                                 
 $\Omega=2\cdot 4^2 \cdot  128$ and $\Omega=2\cdot 8^2 \cdot  128$.             
 In this case the finite size corrections are particularly simple               
\be \label{eq8.12}                                                              
\sigma_\infty=\sigma_\Omega+{1 \over L_{xy}^2 L_t}                              
(c+{3 \over 4} \log L_z -{1 \over 2} \log L_{xy}) \ ,                           
\ee                                                                             
 where $L_x=L_y \equiv L_{xy}$, $L_z$ is the longest extension and $c$          
 is an unknown constant.                                                        
 The obvious advantage of this choice is that practically all mixed             
 configurations contain two surfaces perpendicular to the $z$                   
 direction.                                                                     
 More than two surfaces or surfaces in any other directions are                 
 suppressed by many orders of magnitude.                                        
 (Corrections beyond the gaussian approximation \cite{CASELLE2}                 
 are negligible in our case.)                                                   
                                                                                
\begin{figure}                                                                  
\vspace{9.0cm}                                                                  
\includegraphics{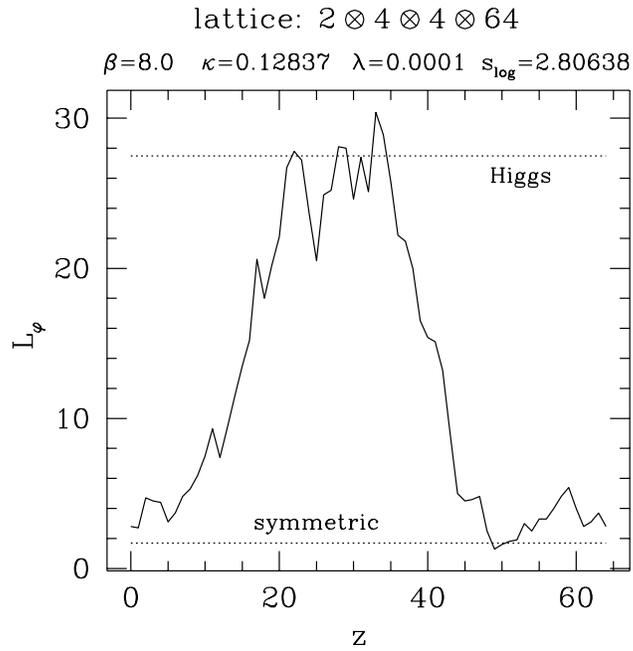}                                           
\begin{center}                                                                  
\parbox{15cm}{\caption{ \label{fig8.1}                                          
 The average values of $L_\varphi$ on $z$-slices in a configuration             
 of $2\cdot 4^2 \cdot  64$ lattice.                                             
 The dotted lines indicate the positions of maxima of the                       
 $L_\varphi$ distributions in the bulk phases.                                                     
}}                                                                              
\end{center}                                                                    
\end{figure}

 The basic picture behind the above formulas assumes                            
 that the two interfaces are thin and do not interact.                          
 Fig.~\ref{fig8.1} is a typical picture of the average $L_\varphi$             
 values for different $z$-slices in a mixed configuration.                      
%%%%%%%%%%%%%%%%%%%%%%%%%%%%%%%%%%%%%%%%%%%%%%%%%%%%%%%%%%%%%%%%%%%%%%%%        
%%%%%%%%%%%%%%%%%%%%%%%%%%%%%%%%%%%%%%%%%%%%%%%%%%%%%%%%%%%%%%%%%%%%%%%%        
%%%%%%%%%%%%%%%%%%%%%%%%%%%%%%%%%%%%%%%%%%%%%%%%%%%%%%%%%%%%%%%%%%%%%%%%        
\begin{figure}                                                                  
\vspace{9.0cm}                                                                  
\includegraphics{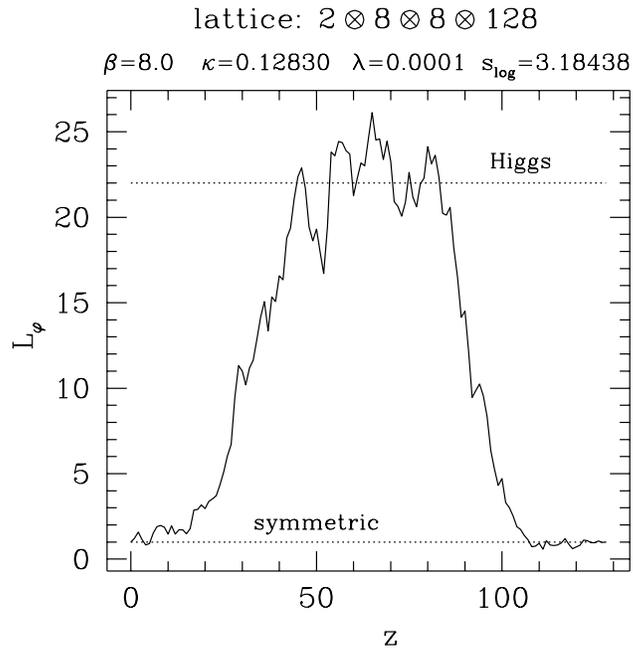}                                          
\begin{center}                                                                  
\parbox{15cm}{\caption{ \label{fig8.2}                                          
 The average values of $L_\varphi$ on $z$-slices in a configuration             
 of $2\cdot 8^2 \cdot  128$ lattice.                                            
}}                                                                              
\end{center}                                                                    
\end{figure}                                                                    
%%%%%%%%%%%%%%%%%%%%%%%%%%%%%%%%%%%%%%%%%%%%%%%%%%%%%%%%%%%%%%%%%%%%%%%%        
 As it can be seen, in the case of $\Omega=2\cdot 4^2 \cdot  64$                
 the interfaces are very close to each other and they cannot                    
 be considered as thin (the spatial width in lattice units is                   
 approximately 30).                                                             
 In this case there is no flat region between the two peaks of the              
 $S_{log}$ distribution.                                                        
 The situation is much better for the even longer lattices 
 (e.g.~$\Omega=2\cdot 8^2 \cdot  128$, see fig.~\ref{fig8.2}).                       
 In the latter case the distribution of the action density $s_{log}$            
 at $\kappa=\kappa_c$ is shown in fig.~\ref{fig4.3}.                           
 It is remarkably flat between the peaks.                                       
 This is another sign that the interfaces are well separated, thus mixed        
 configurations can appear with any bulk phase ratio.                           
                                                                                
 In order to suppress statistical noise we have used a third order              
 polynomial fit to the histograms near the extrema and calculated the           
 extremum values from the fits.                                                 
 (However, due to our good statistics the simple use of the                     
 histogram-bins has resulted in almost the same results.)                       
 Since in case of multicanonical simulations one lattice has been
 attached to each node, we had 128 independent simulations.          
 Our error estimates have been obtained by inspecting these                     
 independent samples.                                                           
                                                                                
 The finite volume interface tensions turn out to be                            
 $\sigma_\Omega/T_c^3=0.63(3)$ for $\Omega=2\cdot 4^2 \cdot  128$ and           
 $\sigma_\Omega/T_c^3=0.80(2)$ for $\Omega=2\cdot 8^2 \cdot  128$,              
 respectively.                                                                  
 Combining these two values one gets from (\ref{eq8.12})                        
\be \label{eq8.13}                                                              
\left( \frac{\sigma_\infty}{T_c^3} \right)^{l2} = 0.83(4) \ .                   
\ee                                                                             
 This result shows that the infinite volume limit is not far                    
 away from our largest volume result and it is                                  
 in very good agreement with the results of the                                 
 previously discussed two-$\kappa$ method in (\ref{eq8.11}).                    
%%%%%%%%%%%%%%%%%%%%%%%%%%%%%%%%%%%%%%%%%%%%%%%%%%%%%%%%%%%%%%%%%%%%%%%%        
                                                                                
\section{Summary and discussion}               \label{sec9}                     
 The main conclusion of this paper  and ref.~\cite{CFHJJM} is that              
 the electroweak phase transition in the SU(2) Higgs model is of                
 first order for Higgs boson masses below 50 $GeV$.                             
 The strongest indication for this is the two-peak structure of                 
 order parameter distributions (see e.g.~fig.~\ref{fig4.3}) showing           
 a strongly suppressed flat region between the peaks which                     
 corresponds to mixed phases.
 For $M_H \simeq 50\; GeV$ the phase transition became rather
 weakly first order.
 Still we could simulate large enough lattices to reach long living
 metastability.                                                   
 In fact, we used this phenomenon to study the properties of both               
 metastable phases separately (see e.g.~section \ref{sec5.2}).                 
 Investigations of the correlation lengths in both phases showed
 that they behave discontinuously at the phase transition.
%%%%%%%%%%%%%%%%%%%%%%%%%%%%%%%%%%%%%%%%%%%%%%%%%%%%%%%%%%%%%%%%%%%%%%%%        
\begin{figure}                                                                  
\vspace{9.0cm}                                                                  
\includegraphics{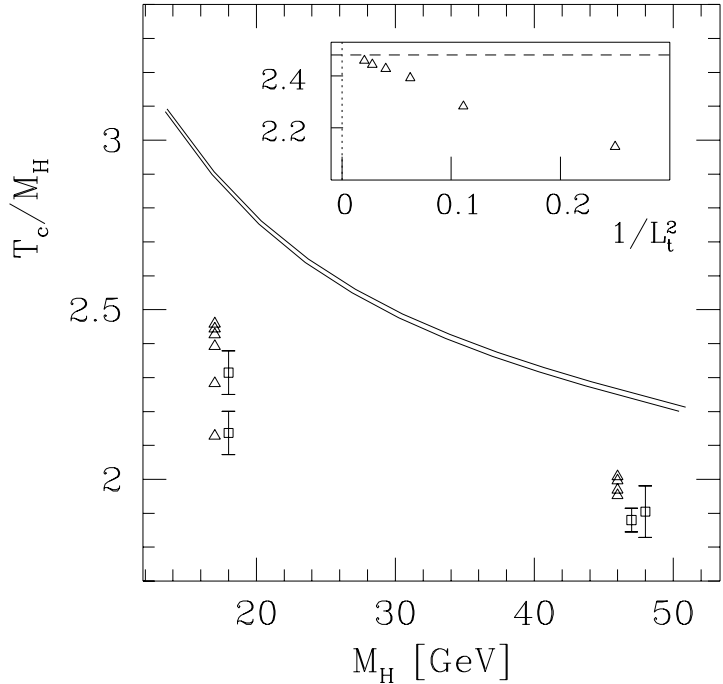}                                          
\begin{center}                                                                  
\parbox{15cm}{\caption{ \label{fig9.1}                                          
 The ratio of the critical temperature and the $T=0$ Higgs mass
 $M_H$ as a function of $M_H$.
 The two curves are the results from the two-loop continuum
 calculation for our renormalized couplings and $M_W \equiv 80\; GeV$.
 The numerical results are the squares with error-bars for
 $M_H \approx 18$ and $49\; GeV$.
 The smaller value in both cases corresponds 
 to $L_t=2$ and the larger one to $L_t=3$.
 The triangles represent the results obtained by the lattice version of 
 the gauge invariant potential which also give growing
 $T_c/M_H$ values.
 This growth is shown by the insert in the upper right corner.
}}                                                                              
\end{center}                                                                    
\end{figure}                                                                    
%%%%%%%%%%%%%%%%%%%%%%%%%%%%%%%%%%%%%%%%%%%%%%                                                                                
 The strength of the first order phase transition can be characterized          
 by the latent heat (section \ref{sec7}) and by the interface tension           
 (section \ref{sec8}).                                                          
 Both these quantities play a central role in the dynamics of first             
 order phase transitions and can be determined in numerical simulations.        
 Our results show that at small Higgs boson masses                              
 ($M_H \simeq 20\; GeV$) the transition is very strong, with a large            
 latent heat (eq.~(\ref{eq7.4})) and interface tension                          
 (eqs.\ (\ref{eq8.11}), (\ref{eq8.13})).                                        
 At Higgs boson masses near $M_H \simeq 50\; GeV$, however, the                 
 phase transition becomes weak: the dimensionless latent heat              
 decreases by about one order of magnitude, the dimensionless interface         
 tension by about two orders of magnitude (see the same equations).             
 The rate of decrease is qualitatively the same as given by two-loop            
 resummed perturbation theory \cite{FODHEB}.                                    
 For this we refer to the figures of ref.~\cite{CFHJJM} which are               
 very similar after the inclusion of the new results here.             
                                                                                
 Since this is our first large scale numerical simulation of the                
 electroweak phase transition, we could not yet achieve a full                  
 quantitative control of the errors.                                           
 This definitely requires further essential work.                               
 Nevertheless, our lattices are large and the statistical errors in most        
 cases small.                                                                   
 In the present paper first attempts were also made to estimate the             
 systematic errors due to finite lattice spacing and finite volume.             
 The comparison of the results on $L_t=2$ and $L_t=3$ lattices                  
 shows remarkable consistency at the 10 \% level.                               
 This means that lattice artifacts are not large.                               
 In particular, the latent heat shows a remarkably good scaling between         
 $L_t=2$ and $L_t=3$ lattices.                                                  
 The apparent general tendency of small scaling violations already at           
 such small temporal lattice extensions is consistent with the                  
 expected dominance of the lowest Matsubara modes motivating the                
 dimensional reduction \cite{KARUSH,JAKAPA,FKRSH}.                                    

 Comparing the results for $T_c$ from the lattice simulations
 and those from perturbation theory, the values 
 obtained from the simulations are always smaller.
 However, the proper comparison should be done not with
 continuum but with lattice perturbation theory, for 
 relatively small $L_t$'s.
 We have carried out this analysis (see fig.~\ref{fig9.1}). 
 In order to determine the critical points and masses 
 the gauge invariant effective potential of section \ref{sec3}  
 has been used (again, the role of the symmetric phase and effects
 due to higher order corrections were neglected).
 One can see that for $m_H=18\ GeV$ the perturbative result
 is in complete agreement with the lattice one.
 A scaling violation can be observed and it goes
 approximately like $1/L_t^2 \propto a^2$.
 (See the inlet of fig.~\ref{fig9.1}, where $L_t \approx 5-7$ practically 
 reaches $2.47$, indicated as a dashed line.
 This value corresponds to  $L_t\rightarrow \infty$, which can be
 also obtained by the use of the one-loop effective potential
 without high-temperature expansion.)
 In the case of $m_H \approx 49\ GeV$ the
 situation is somewhat different.
 The perturbative values are larger than those obtained by the
 lattice simulations.
 The scaling violation is much smaller and within the 
 errors it is reproduced by the perturbative treatment.                                      

 Up to now we did only a few checks of finite volume effects.                   
 These showed that the results are only changing at the few percent             
 level.                                                                         
 Further systematic tests have to be done in the future.                        
 Nevertheless, the lattices are large enough to support strong                  
 metastability of the two phases.                                               
 Therefore the conclusions concerning the strength of the first order           
 phase transitions are firm. 
                                                                                
 In general, numerical simulations of the electroweak phase transition          
 in the four dimensional SU(2) Higgs model in the range of Higgs boson          
 masses below $M_H \simeq 50\; GeV$ turned out to be feasible and               
 powerful.                                                                      
 For these relatively light Higgs masses there is no need to go to              
 the three dimensional reduced model.                                           
 The situation could, however, be different for heavier Higgs bosons.           
 In fact, recent numerical studies of the reduced SU(2) Higgs                   
 model are concentrating on $M_H \simeq M_W$ \cite{FAKARUSH}.                   
                                                                                
 In future numerical simulations in the four dimensional model one              
 has to consider also $L_t=4$ lattices for confirming the smallness of          
 scaling violations.                                                            
 A systematic study of finite volume effects is desirable.                      
 Of course, an extension towards heavier Higgs bosons would be very             
 interesting.                                                                   
                                                                                
%%%%%%%%%%%%%%%%%%%%%%%%%%%%%%%%%%%%%%%%%%%%%%%%%%%%%%%%%%%%%%%%%%%%%%%%        
                                                                                
\vspace{1cm}                                                                    
{\large\bf Acknowledgements}                                                    
\vspace{3pt}                                                                    
                                                                                
\noindent                                                                       
We thank W. Buchm\"uller,  F. Csikor, A. Hebecker, M. L\"uscher,                
R. Sommer and Ch. Wetterich for essential comments and proposals during         
the course of this work.                                                        
In particular, we  would like to acknowledge the participation of               
F. Csikor in parts of this work related to the multicanonical                   
simulations.                                                                    
Z. F. was partially supported by Hung. Sci. Grant under                         
Contract No. OTKA-F1041/3-2190.

%%%%%%%%%%%%%%%%%%%%%%%%%%%%%%%%%%%%%%%%%%%%%%%%%%%%%%%%%%%%%%%%%%%%%%%         

%%%%%%%%%%%%%%%%%%%%%%%%%%%%%%%%%%%%%%%%%%%%%%%%%%%%%%%%%%%%%%%%%%%%%%%%        
\newpage \appendix                                                              
\section{Appendix: multicanonical heatbath algorithms}                          
                                                                                
 Here we describe the modifications for the heatbath algorithms in              
 multicanonical simulations.
 For sake of simplicity we first explain the general strategy in case
 of the gauge field and then the realization of the algorithms
 for the fields $U$ and $\varphi$.                  
 We use the notations introduced in section \ref{sec2.3}.                       
 In general, the indices {\em old} and {\em new} will denote the                
 value before the update and the proposal, respectively.                        
                                                                                
 We distinguish two cases.                                                      
 If all possible values of the new action $S_{log}^{new}$                       
 remain in the interval $I_k=(S^k,S^{k+1}]$, one has to replace                 
 $S$ by $(1+\beta_k)S$ and the generation of the distribution is as             
 usual (see section \ref{sec2.1}).                                              
 If $S_{log}^{new}$ crosses an interval boundary, the two distributions         
 which are different for the two intervals have to be matched.                  
 This is done in such a way that for the interval $I_k$ the                     
 distribution becomes $W_k\sim \exp[-(1+\beta_k)S]$                             
 and for $I_{k+1}$ it is $W_{k+1}\sim \exp[-(1+\beta_{k+1})S]$.                 
 In addition the distribution for the combined interval $I_k$ and               
 $I_{k+1}$ have to be continuous.                                               
 The idea is to construct a distribution which is a hull of $W_k$               
 and $W_{k+1}$.                                                                 
 To obtain the desired distribution in each interval one has to                 
 correct the hull distribution by imposing an accept-reject step.               
 The correction factor (denoted by $\Theta$) has to be positive and not         
 larger than one.                                                               
 Clearly if the hull distribution is close to $W_k$ and $W_{k+1}$               
 the correction factor is about one and the acceptance rate is high.            
                                                                                
%%%%%%%%%%%%%%%%%%%%%%%%%%%%%%%%%%%%%%%%%%%%%%%%%%%%%%%%%%%%%%%%%%%%%%%%        
\subsection{Gauge field}                                                        
\begin{figure}                                                                  
\vspace{9.0cm}                                                                  
\includegraphics{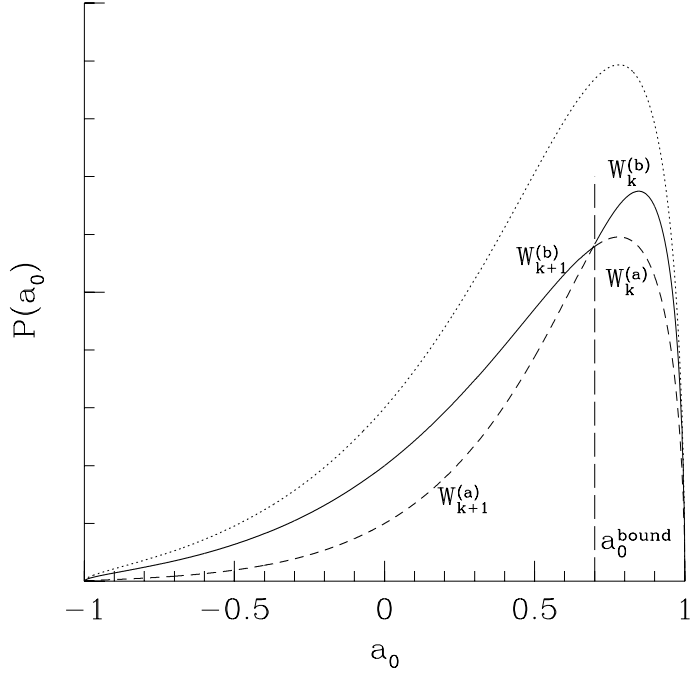}                                               
\begin{center}                                                                  
\parbox{15cm}{\caption{\label{figapp.1}                                         
 The figure illustrates the different cases for gauge field updating.           
 In case $\beta_k < \beta_{k+1}$ the desired multicanonical                     
 distribution is represented by the dashed line, in case                        
 $\beta_k > \beta_{k+1}$ by the continuous line.                                
 In the first case one generates the distribution $W_k^{(a)}$ for the           
 whole interval $-1 \leq a_0 \leq 1$, the continuation of which is the          
 solid line for $a_0 < a_0^{bound}$, and corrects only                          
 if $a_0 < a_0^{bound}$.                                                        
 In the second case one generates a hull distribution                           
 (dotted line) proportional to  $W_{k+1}^{(b)}$ in the whole interval.          
}}                                                                              
\end{center}                                                                    
\end{figure}

 Let's first discuss the gauge updating.                                        
 For the standard heatbath algorithm the main task is to generate a             
 distribution                                                                   
\begin{eqnarray}                                                                
\label{eqnapp.1}                                                                
P(a_0) da_0 \sim (1-{a_0}^2)^{\frac{1}{2}} \exp(\xi a_0) da_0 &                 
\hspace{2em} (-1 \leq a_0 \leq 1,\; \xi >0) \ .                                 
\end{eqnarray}                                                                  
 This is done by the algorithm described in \cite{KENPEN}.                      
 In the multicanonical case the first step is to check, as                      
 discussed above, whether the global action                                      
 $S_{log}^{new}$ can cross an interval boundary.                                
 This depends on $a_0$ through                                                  
 $S_{log}^{new}(a_0^{new})=S_{log}^{old}+\xi(a_0^{old}-a_0^{new})$.             
 If the minimal action $S_{log}^{new}(1)$ and the maximal action                
 $S_{log}^{new}(-1)$ are in the same interval $I_k$ one                         
 replaces $\xi$ by $\xi(1+\beta_k)$ in (\ref{eqnapp.1})                         
 and proceeds in the usual way.                                                 
 In case that the action crosses an interval boundary $S^{k+1}$,                
 which defines the corresponding $a_0^{bound}$ through                          
 $S^{k+1}=S_{log}^{new}(a_0^{bound})$, one constructs the distribution          
 in the following way:                                                          
\begin{enumerate}                                                               
\item[(a)]  \large$\beta_k < \beta_{k+1}$\normalsize     \\                     
Generate the hull distribution                                                  
\be                                                                             
P(a_0) da_0 \sim (1-{a_0}^2)^{\frac{1}{2}}                                      
\exp[\xi (1+\beta_k) a_0] da_0 \ .                                              
\ee                                                                             
 If $a_0\geq a_0^{bound}$ ($S_{log}^{new}\in I_k$) no correction factor         
 is needed.                                                                     
 For $a_0<a_0^{bound}$ ($S_{log}^{new}\in I_{k+1}$) one has to impose           
\be                                                                             
\Theta = \exp[\xi(\beta_{k+1}-\beta_k)(a_0-a_0^{bound})]                        
\ee                                                                             
 as an additional accept-reject factor (see figure \ref{figapp.1}).             
\item[(b)] \large$\beta_k > \beta_{k+1}$\normalsize     \\                      
 Generate the hull distribution                                                 
\be                                                                             
P(a_0) da_0 \sim (1-{a_0}^2)^{\frac{1}{2}}                                      
\exp[\xi (1+\beta_{k+1}) a_0] da_0                                              
\ee                                                                             
 and correct for the following factor (see figure \ref{figapp.1}):              
\be                                                                             
\Theta= \left\{                                                                 
\begin{array}{lll}                                                              
\exp[\xi(\beta_k-\beta_{k+1})(a_0-1)] & \hspace{2em}                            
a_0 \geq a_0^{bound}&(S_{log}^{new}\in I_k) \ ,         \\[1em]                 
\exp[\xi(\beta_k-\beta_{k+1})(a_0^{bound}-1)]&\hspace{2em} a_0 <                
a_0^{bound}&(S_{log}^{new}\in I_{k+1}) \ .              \\                      
\end{array}                                                                     
\right.                                                                         
\ee                                                                             
\end{enumerate}                                                                 
 To avoid that the action $S_{log}^{new}$ crosses more                          
 than one interval boundary, the intervals must be large enough.                
                                                                                
 In practice the values of $\beta_k$'s are small                                
 (${\cal O}(10^{-2}-10^{-3})$).                                                 
 Therefore the two curves are closer together than in figure                    
 \ref{figapp.1} and the acceptance rate is high.                                
 Also the values of $\xi$ are higher, i.e.\ there is a sharp peak near          
 $a_0=1$.                                                                       
%%%%%%%%%%%%%%%%%%%%%%%%%%%%%%%%%%%%%%%%%%%%%%%%%%%%%%%%%%%%%%%%%%%%%%%%        
%%%%%%%%%%%%%%%%%%%%%%%%%%%%%%%%%%%%%%%%%%%%%%%%%%%%%%%%%%%%%%%%%%%%%%%%        
                                                                                
%%%%%%%%%%%%%%%%%%%%%%%%%%%%%%%%%%%%%%%%%%%%%%%%%%%%%%%%%%%%%%%%%%%%%%%%        
\subsection{Scalar field}                                                       
 The algorithm which is used for $\varphi$-updating was described               
 in section \ref{sec2.1}.                                                       
 We write the action of eq.~(\ref{zetaaction}) as a sum of the                 
 quadratic and quartic part and a constant                                      
\be                                                                             
S(\varphi_x)=S_{(2)}(\varphi_x)+S_{(4)}(\varphi_x)+const.'\hspace{1em}.         
\ee                                                                             
 For the multicanonical simulations one has to generate the                     
 distribution                                                                   
\begin{eqnarray}                                                                
P(\varphi_x) d^4\varphi_x \sim                                                  
\exp[-S(\varphi_x) - \beta_k S_{log} - \alpha_k]                  
d^4\varphi_x                                                                    
\end{eqnarray}                                                                  
 for the four components of $\varphi_x$.                                        
 The first step is to generate the distribution                                 
\be                                                                             
P(\varphi_x) d^4\varphi_x \sim                                                  
\exp[-(1+\beta_{min}) S_{(2)}(\varphi_x)] d^4\varphi_x                          
\ee                                                                             
 as a hull analytically by replacing                                            
 $S_{(2)}(\varphi_x)\rightarrow (1+\beta_{min})S_{(2)}(\varphi_x)$.             
 $\beta_{min}$ denotes the minimal $\beta_k$.                                   
 Let $I_k$ denote the interval in which the minimum $S^0_{log}$                 
 of the action can be found.                                                    
 If $S_{log}^{new}$ is in this interval $I_k$, the correction factor is         
\be                                                                             
\Theta_1=\exp[(\beta_{min}-\beta_k)S_{(2)}                                      
(\varphi_x)+ 3 \beta_k \log(\rho_x)                                             
-(1+\beta_k)S_{(4)}(\varphi_x) -c]                                              
\ee                                                                             
 where the meaning of the constant $c$ is described below.                      
 In case of crossing an interval border, i.e.\                                  
 $S_{log}^{new} = S-3\sum_x \log(\rho_x)>S^{k+1}$, one needs an                 
 additional factor                                                              
\be                                                                             
\Theta_2=\exp\{[S_{log}^{new} - S^{k+1}](\beta_k-\beta_{k+1})\} \ .             
\ee                                                                             
 The total correction factor is in this case $\Theta_1 \Theta_2$.               
 This procedure generates the action distribution in $I_k$ and in               
 $I_{k+1}$ correctly.                                                           
 Due to the definition of $I_k$, $S_{log}^{new}$                                
 cannot be in $I_j$ ($j<k$).                                                    
 Since our intervals are large enough the probability that                      
 $S_{log}^{new}$ is in $I_j$ ($j>k+1$) is negligible.                           
 In practice it is quite difficult to find  $S^0_{log}$, but a slightly         
 smaller $\bar{S}_{log}<S^0_{log}$ can be used, which could lead to a           
 somewhat smaller acceptance rate.                                              
 In addition it is necessary that the whole correction factor (which            
 includes all terms except the analytically generated quadratic term)           
 has to be smaller or equal one.                                                
 To be sure that no problem arises we subtracted a small positive               
 extra term $c$ in the exponent of $\Theta_1$.                                  
 This term reduces the acceptance.                                              
 It must be tuned in such a way that the correction factor is never             
 greater than one and the acceptance rate is high.                              
                                                                                
 Of course, there is always only one accept-reject step in both                 
 updating algorithms.                                                           
 This means that the analytically generated distribution                        
 will be corrected with the product of all factors.                             
                                                                                
%%%%%%%%%%%%%%%%%%%%%%%%%%%%%%%%%%%%%%%%%%%%%%%%%%%%%%%%%%%%%%%%%%%%%%%%        
                                                                                
\end{document}